# Physics of Meteor Generated Shock Waves in the Earth's Atmosphere – A Review

Elizabeth A. Silber[1*], Mark Boslough[2], Wayne K. Hocking[3], Maria Gritsevich[4,5,6], Rodney W. Whitaker[7]

[1]Department of Earth, Environmental and Planetary Sciences, Brown University, Providence, RI, 02912, USA; [2]Department of Earth and Planetary Sciences, University of New Mexico, Albuquerque, NM, 87131, USA; [3]Department of Physics and Astronomy, University of Western Ontario, London, Ontario, N6A 3K7, Canada; [4]Department of Physics, P.O. Box 64, FI-00014 University of Helsinki, Finland; [5]Institute of Physics and Technology, Ural Federal University, Ekaterinburg, Russia; [6]Dorodnicyn Computing Centre, Russian Academy of Sciences, Moscow, Russia; [7]Los Alamos National Laboratory, EES-17 MS F665, PO Box 1663 Los Alamos, NM, 87545, USA

**Abstract**

Shock waves and the associated phenomena generated by strongly ablating meteoroids with sizes greater than a few millimeters in the lower transitional flow regime of the Earth's atmosphere are the least explored aspect of meteor science. In this paper, we present a comprehensive review of literature covering meteor generated shock wave phenomena, from the aspect of both meteor science and hypersonic gas dynamics. The primary emphasis of this review is placed on the mechanisms and dynamics of the meteor shock waves. We discuss key aspects of both shock generation and propagation, including the great importance of the hydrodynamic shielding that develops around the meteoroid. In addition to this in-depth review, the discussion is extended to an overview of meteoroid fragmentation, followed by airburst type events associated with large, deep penetrating meteoroids. This class of objects has a significant potential to cause extensive material damage and even human casualties on the ground, and as such is of great interest to the planetary defense community. To date, no comprehensive model exists that accurately describes the flow field and shock wave formation of a strongly ablating meteoroid in the non-continuum flow regime. Thus, we briefly present the current state of numerical models that describe the comparatively slower flow of air over non-ablating bodies in the rarefied regime. In respect to the elusive nature of meteor generated shock wave detection, we also discuss relevant aspects and applications of meteor radar and infrasound studies as tools that can be utilized to study meteor shock waves and related phenomena. In particular, infrasound data can provide energy release estimates of meteoroids entering the Earth's atmosphere. We conclude with a summary of unresolved questions in the domain of meteor generated shock waves; topics which should be a focus of future investigations in the field.

[*]Corresponding Author: Elizabeth A. Silber, Department of Earth, Environmental and Planetary Sciences, Brown University, 324 Brook St., Box 1946, Providence, RI, 02912, USA, E-mail: elizabeth_silber@brown.edu

# 1. Introduction

According to the definition by Thomas Kuhn (Beech, 1988), meteor science has reached maturity, yet by some other arguments may be a science in decline (Gilbert, 1977). However, these remarks do not necessarily apply to all sub-disciplines of meteor science. There are still a number of major questions that remain unanswered, especially in regard to nonlinear and short time energetic processes that take place during the meteoroid ablation, such as shock waves. Historical examples of extraterrestrial impacts, such as the Tunguska (e.g., Chyba et al., 1993) and the more recent Chelyabinsk (Brown et al., 2013; Popova et al., 2013) events, serve as a reminder of the destructive potential of shock waves generated by large meteoroids, and the importance of studies of shock wave phenomena.

Detailed studies in the field of meteor physics that consider theoretical, observational and computational studies of the dynamics and effects of meteor shock waves are relatively sparse. That is not to say that there is a lack of understanding of rarefied gas dynamics and hypersonic flow (Brun, 2009; Anderson, 2006; Cercignani, 2000); rather, that area of research has been, and still is, confined and maintained within the domain of specialized applications of re-entry studies (Sarma, 2000), there being little or no cross disciplinary work with investigators in the meteor field. Thus, the main motive behind this review is to consolidate and bring the state of the field up to date and to facilitate further understanding of and inquiry into those unresolved questions in meteor science, in particular about meteor generated shock waves and related phenomena which in extreme and rare cases can present considerable danger to civilian lives and infrastructure (e.g., Brown et al., 2013; Popova et al., 2013; Pelton et al., 2015; Kulagin et al., 2016; Tapia et al., 2017). We present a long needed condensed survey of the available (and sparse) literature covering fundamental and up-to-date aspects of meteor related shock wave phenomena and attempt to balance it with a review of the related interdisciplinary work from hypersonic gas dynamics (and other related disciplines). The primary emphasis of this review is placed on the cylindrical shock waves generated by meteoroids with diameters greater than four millimeters and up to large fireballs. Additionally, the dynamics and hazards associated with large deep penetrating meteoroids (e.g., Boslough et al., 1995; Silber et al., 2009; Brown et al., 2013; Popova et al., 2013) are also presented in some depth.

This review paper is organized as follows: Section 2 presents a general overview of meteoroid entry in the atmosphere; Section 3 describes the Knudsen number and flow regimes of meteoroids; Section 4 presents a detailed overview of hydrodynamic shielding and implications to the formation of shock waves; Section 5 describes transition from hydrodynamic shielding to shock wave; and Section 6 provides a detailed breakdown and "anatomy" of meteor generated shock waves. While the review of meteor generated shock waves technically concludes with Section 6, we add three additional sections to provide the reader with further insight into applications and implications of the meteor generated shock waves. These sections include analytical and numerical approaches (Section 7), meteoroid fragmentation (Section 8), airbursts and NEO threat (Section 9), and radar and infrasound observations of meteor generated shock waves (Section 10). We conclude this review paper with Section 11, where we present some remaining unresolved problems in the field of meteor shock wave physics.

# 2. Meteoroids in the Atmosphere

The scientific inquiry into the nature of meteor phenomena can be traced to the 19$^{th}$ century (Hughes, 1982), although meteor science experienced a renaissance period after the Second World War, following the rapid advancements in radar and spectroscopic techniques. Moreover, with the advent of the space-race, early studies were also motivated by problems of ascending and re-entry vehicles, and protection of space assets (DeVincenzi et al., 1998; Ben-Dor et al., 2000). In general, the study of meteor phenomena is of great importance beyond narrow sub-disciplinary studies. The interplanetary and interstellar particles may have brought chemicals necessary for seeding life on Earth (Plane, 2012). Meteoroids are also considered to be a source of the metallic layer in the mesosphere and lower thermosphere (MLT) above 80 km and play an important role in the chemistry of the upper atmosphere (Ceplecha et al., 1998; Plane, 2012; Plane et al., 2015). However, it is interesting to note that, while meteor science has branched into many sub disciplines, such as the study of upper atmospheric density, temperature, scale height (e.g., Manning et al., 1959; Tsutsumi et al., 1994; Marsh et al., 2001), diffusion coefficients, wind velocity and turbulence (e.g., Hocking et al., 2001; Jones et al., 2003; Hocking et al., 2016a, 2016b), meteor generated shock waves have received the least attention in the literature. Before we engage in a discussion about the shock waves, we shall briefly revisit the fundamentals of the physics of meteor phenomena.

Some estimates suggest that around $10^5$ tons/year of interplanetary and space dust and particles (collectively called meteoroids) of mainly chondritic origin (Plane, 2012) impact the Earth's atmosphere with a peak in the size (diameter) and mass distribution of around $2 \cdot 10^{-4}$ m and 10 μg, respectively (Kalashnikova et al., 2000; Plane, 2012). While the combined number of particles entering the Earth's atmosphere daily is about $10^8$ (Barri, 2010a) (or about 300 tons), only a small fraction of that number corresponds to cm sized and larger meteoroids (Moorhead et al., 2017). The total mass influx has been slightly revised recently; however, there are still large uncertainties in those estimates (Drolshagen et al., 2017; Moorhead et al., 2017). Entry (pre-atmospheric) particle velocities range from 11.2 km/s to 72.5 km/s (Baggaley, 1980; Baggaley, 2002) with mean values between 25 - 30 km/s determined through radar observations (Janches et al., 2006) and 20.6 km/s for Near-Earth Objects (NEO) estimated through modeling (Greenstreet et al., 2012).

Upon entering the upper Earth atmosphere, a meteoroid experiences an initial mass loss by sputtering (Hill et al., 2004; Rogers et al., 2005; Vinković, 2007) as a result of high energy collisions with a rarefied atmospheric gas molecules. At lower altitudes, the meteoroid encounters the exponentially increasing atmospheric density and undergoes rapid heating and subsequent differential ablation (Vondrak et al., 2008). However, ablation takes place only if the particle can reach sufficiently high temperatures (Plane, 2012). As micrometeoroids are susceptible to rapid heat loss by thermal radiation (Popova et al., 2001), fusion temperatures are reached by meteoroids with masses exceeding $10^{-7}$ g and a diameter greater than ~ $10^{-6}$ m (Plane, 2012). Besides fragmentation, additional factors such as material properties and particle velocity play a critical role in onset, efficiency and type of ablation (Bronshten, 1983; Campbell-Brown et al., 2004; Popova, 2005). Particles in the size range 0.1 – 10 cm ablate in the region of the atmosphere between approximately 70 and 100 km altitude. Larger meteoroids (including meter-sized ones) penetrate much deeper in the lower atmosphere where they deposit most of their energy (e.g., Popova, 2005; Gritsevich, 2008; Gritsevich et al., 2017). In principle, the ablation plays an essential role for the meteoroids in the range of roughly $10^{-6}$–100 m in diameter (Popova, 2005). For an ablating meteoroid, the energy required to completely vaporize the particle is several orders of magnitude smaller than the initial kinetic energy imparted by collisions with the atmospheric molecules (Romig, 1965; Zinn et al., 2004). A review of ablation and ablation models is given by Popova (2005). While ablation models (Campbell-Brown and Koschny, 2004) are capable of reasonably reproducing the ablation of faint meteoroids in the size range of $1 \cdot 10^{-4}$ to $2 \cdot 10^{-3}$ m, a more recent study (Campbell-Brown et al., 2013) discusses the shortcomings of numerical approaches to meteoroid ablation and demonstrates the need for significant improvement in meteoroid ablation models. In the context of differential ablation, it is important to note that early results of the experimental heating of meteoric material (Notsu et al., 1978) and the thermodynamic modeling of vaporization of chondritic magma (Fegley et al., 1987) supported the hypothesis of differential ablation, as it was indicated that the meteoric elements are released into the ambient atmosphere according to their differing volatilities. More recent models of differential ablation were successful in resolving the nature of the distinct metallic layer in the MLT (e.g., McNeil et al., 1998; Vondrak et al., 2008; Plane, 2004, 2012; Feng et al., 2013; Marsh et al., 2013; Dawkins et al., 2015; Plane et al., 2015; Bones et al., 2016). However, if models of differential ablation are based on oversimplified assumptions, then the potential uncertainties may arise if such models neglect the fact that a significant amount of volatile species might be bound in more thermodynamically stable oxide phases.

The optically detectable luminous phenomenon (applicable to sufficiently large particles), resulting from de-excitation of ablated and collisionally excited molecules and atoms, is called a meteor. In principle, the radiative stage of the meteor evolution is a result of ablated and ionized plasma radiative energy loss that takes place during the collisional deceleration. The ablated meteoric atoms and ions, and the initially entrained and modified atmospheric gas decelerate to the ambient kinetic velocities within several hundred meters (Jenniskens et al., 2004a). If a fragment survives the passage through the atmosphere and lands on the ground, it is called a meteorite. We will not discuss these objects further.

Several different mechanisms are involved in the meteoroid ablative mass loss; the prevalence of specific ablation mechanisms is a function of meteoroid mass, velocity, composition and altitude (Bronshten, 1983). High energy collisions with the local atmosphere involve the exchange of translational, rotational, and vibrational energy and lead to excitation, dissociation and ionization of both incident atmospheric molecules and ejected meteoric atoms (Bauer, 1990; Dressler et al., 2001; Zinn et al., 2004; Panesi et al., 2011). Depending on the meteoroid velocity, the kinetic energies imparted collisionally to ablated meteor atoms may exceed several hundred electron volts (eV), while the initial kinetic energy of any free electrons may approach 10 eV (Baggaley, 1980; Hocking et al., 2016b). Those high energy collisions that lead to the potential formation of new species are generally referred to as reactive collisions (for a detailed discussion about reactive flows see Brun, 2009).

The density of ablated meteoric atoms and molecules is much greater than the density of the incident local atmosphere. The simple picture is that the high temperature ablated meteoric atoms (which are ionized by collisions or by radiative emissions) initially expand adiabatically in the meteor wake and form a cylindrical volume of ionized plasma with some explosively formed initial radius $r_0$ (Jones, 1995; Jones et al., 2005; Hocking et al., 2016b). We consider the meteor wake to simply be the region immediately aft of the meteoroid. It is also the region that approximately corresponds to the meteor radiative stage (or the stage when ablated vapor is moving and it has not been slowed down to ambient gas velocities) (McCrosky, 1958; Bronshten, 1983). Upon reaching the initial radius (Jones, 1995), this dynamically stable volume of plasma vapor is in pressure (but not in temperature) equilibrium with the local atmosphere and is typically several kilometers long (e.g., see Lees et al., 1961, 1962; Jones, 1995; Jones and Campbell-Brown, 2005; Hocking et al., 2016b). This is a generally reasonable interpretation for the formation of the trail with some initial radius of ablating micrometeoroids. However, for a category of larger meteoroids ($d_m \geq 4$ mm), the formation of shock waves (resulting in heating of ambient atmosphere) and higher residual temperatures in the meteor wake need to be considered during the initial stage of meteor trail postadiabatic expansion (Hocking et al., 2016b).

Higher temperatures in the initial meteor train may be due to the phenomena of "frozen" and non-equilibrium flows of high energy ablated meteoric ions and atoms (to be discussed later) mixed with energized dissociated (and ionized) atmospheric constituents that were swept behind the flow field. This means that the postadiabatic train may not be in initial thermal equilibrium with the atmosphere at the time of its formation (Jenniskens et al., 2004a; Zinn et al., 2005). Indeed, this may indicate that the initial radial expansion of plasma (following the adiabatic trail formation) exceeds the rate of ambipolar diffusion (see Hocking et al., 2016b).

During the trail formation, the radial expansion of ablated high temperature material may also be complicated by the effect of local turbulent diffusion, driven by the flow field velocity, temperature and density gradients (Lees and Hromas, 1961, 1962; Silber et al., 2017a). The degree of ionization in the meteor train depends on the so called ionization coefficient (McKinley, 1961; Jones, 1997; Jones et al., 2001; Weryk et al., 2013). The ionized meteor train, as an artifact of meteor passage through the atmosphere, can be either observed visually or detected by radar (Hocking et al., 2001; 2016b), as the quasineutral plasma expands into the ambient atmosphere under height dependent ambipolar diffusion (Francey, 1963; Holway, 1965; Galligan et al., 2004; Oppenheim et al., 2015). Some aspects of the short lasting hyperthermal chemistry which take place on the boundaries of high temperature cylindrical meteor trains in the initial stages of postadiabatic expansion have been studied by Silber et al. (2017a). Thermally driven chemical processes within the meteor wake and high temperature trains were discussed by Berezhnoy et al. (2010) and in earlier work by Menees et al. (1976) and Park et al. (1978).

Contrary to still unresolved questions about thermally driven chemical reactions and uncertainty in thermalization times of the meteor train reported in the study of Jenniskens et al. (2004a), the aspects of thermalized meteor train chemistry are well understood (e.g., Baggaley, 1978, 1979; Plane, 2003, 2012; Whalley et al., 2011; Plane et al., 2015). Further comprehensive reviews and discussions of the meteoroid ablation process can be found in Bronshten (1983), Ceplecha et al. (1998), Plane (2012) and computational work by Boyd (2000), Popova et al. (2001), Zinn et al. (2004) and Zinn and Drummond (2005). Similarly, a review of meteoroid ablation models is given by Popova (2005).

The classical energy balance equation for a spherical meteoroid under the assumption of an isotropic energy flux can be expressed as:

$$\frac{1}{2}\pi r_m^2 \Lambda \rho_a v^3 = 4\pi r_m^2 \varepsilon \sigma_{SB}(T_s^4 - T_0^4) + Q\frac{dm}{dt} + \frac{4}{3}\pi r_m^3 \rho_m C \frac{dT_m}{dt} \qquad (1).$$

Here, $r_m$ is the meteoroid radius, $\Lambda$ is the dimensionless heat transfer coefficient, which is a measure of efficiency of the collision process in converting kinetic energy into heat (McKinley, 1961), and $Q$ is the latent heat of vaporization. The specific heat and density of the meteoroid along with density of air and meteoroid velocity are denoted by $C$, $\rho_m$, $\rho_a$ and $v$, respectively. The Stefan-Boltzmann constant is classically defined by $\sigma_{SB}$, and $\varepsilon$ is the emissivity of the meteoroid. $T_s$, $T_0$, $T_m$, are the temperature of the meteoroid surface, temperature of the ambient air and some mean temperature of the meteoroid at some small time interval, respectively (e.g., see Popova, 2005; Plane, 2012). The terms on the left hand side (LHS) describe the energy input per area by colliding air molecules. On the right hand side (RHS), the first term is the radiative loss and the second term describes the heat consumed in the transfer of particle mass into the gas phase (or vaporization). The last term on the right represents the energy losses due to heat conduction (i.e., phase transitions and heating).

For the purpose of completeness, let us list the rest of the basic equations that govern meteoroid motion in the atmosphere. We will limit the exposition to single body ablation and discrete, gross fragmentation (Ceplecha et al., 1998). Consider a meteoroid with mass ($m$), radius ($r_m$), density ($\rho_m$) and projected cross-sectional area ($S$), moving at velocity ($v$) and having a drag coefficient ($C_D$). The meteoroid then sweeps through a volume of air in a time increment ($dt$), transferring kinetic energy to the atmosphere. Now we can introduce the dimensionless shape factor ($A_s$), which relates the mid-sectional area of the meteoroid to its volume ($V_m$) through: $A_s = \frac{S}{V_m^{2/3}}$. Then, in general, the cross-sectional area can be expressed in terms of the meteoroid density and meteoroid mass:

$$S = A_s \left(\frac{m}{\rho_m}\right)^{2/3} \qquad (2).$$

The value of $A_s$ will depend on the shape of the meteoroid. For example, for the sphere, $A_s = 1.21$; for a hemisphere, $A_s = 1.92$; and for a cube moving face on to the flow, $A_s = 1$ (McKinley, 1961). However, since meteoroids tumble during their flight through the atmosphere, it is generally assumed that irregularly shaped bodies will have their shape factor approach that of a sphere. In cases where a more detailed knowledge of the meteoroid shape as a function of rotation rate (Levin, 1956) is needed, the light curve analysis from optical observations can be used to better constrain this information (e.g., Gritsevich et al., 2011; Bouquet et al., 2014 and references therein).

As the meteoroid propagates through the atmosphere, it sweeps out a volume of air of size *Svdt* in time *dt* The mass of the volume can be represented in terms of the atmospheric density ($\rho_a$) as $dm_a = \rho_a Svdt$. Substituting for *S* from Eq. (2), one can arrive to the expression for the rate of changing air mass ($m_a$) in unit time encountered by the meteoroid (McKinley, 1961) as:

$$dm_a = \rho_a \left(\frac{m}{\rho_m}\right)^{2/3} A_s \, vdt \qquad (3).$$

As a result of impacts with the meteoroid, this parcel of air with a height-specific mass density will transfer momentum to the meteoroid. This in turn results in a rate of change of momentum of the meteoroid:

$$\frac{d(mv)}{dt} = v\frac{dm}{dt} + m\frac{dv}{dt} \qquad (4).$$

If the meteoroid does not ablate very fast, i.e., *m*≫*dm*, the term *vdm/dt* in Eq. (4) is negligible and therefore ignored. On the other hand, the air particles in this volume will gain momentum per unit time:

$$\frac{C_D}{2}\frac{dm_a}{dt}v = \frac{C_D}{2} A_s \left(\frac{m}{\rho_m}\right)^{\frac{2}{3}} \rho_a v^2 \qquad (5).$$

The drag equation can be derived by equating the loss of momentum per second of the meteoroid (*mdv/dt*) with the momentum gained by the air particles:

$$\frac{dv}{dt} = -\frac{C_D A_s \rho_a v^2}{2m^{1/3}\rho_m^{2/3}} \qquad (6).$$

The negative sign in Eq. (6) indicates deceleration. The rate of mass loss is proportional to the kinetic energy of the hypersonic flow impinging on the meteoroid. This can be represented with the differential mass equation (McKinley, 1961), also known as the mass-loss equation (Ceplecha et al., 1998):

$$\frac{dm}{dt} = -\frac{\Lambda A_s \rho_a v^3 m^{2/3}}{2Q\rho_m^{2/3}} \qquad (7).$$

Here, *Q* is the heat of ablation of the meteoroid material (or energy required to ablate a unit mass of the meteoroid).

Our goal here is to clearly develop and present the concepts related to meteor generated shock waves, something that has not been frequently done in literature. Furthermore, for the purpose of brevity and the expository nature of this review, the use of mathematical (and numerical) tools in analysis of meteor related parameters and subsequent shock wave phenomena is kept at a minimum, unless explicitly needed for the completeness of the discussion. Most of the mathematical concepts needed to study shock waves are comprehensively discussed, derived and presented in excellent monographs (e.g., Hayes et al., 1959; Bronshten, 1983; Zel'dovich et al., 2002; Anderson, 2003, 2005, 2006) and when necessary, are suggested for further reading. We now turn attention to the characterization of flow regimes and discussion about why their classical definition may not always be applicable to meteoroids in the atmosphere.

## 3. Knudsen Number and Flow Regimes

In hypersonic gas dynamics, the Knudsen number (*Kn*) is one of the most important parameters when assessing shock wave formation (Anderson, 2006). The Knudsen number is the similarity parameter that characterizes the type of hypersonic flow; it is defined as the ratio of the local atmospheric mean free path to the characteristic dimension of the body (in our case, the meteoroid radius) (Silber et al., 2017a). There are four basic types of flow regimes that can be distinguished (Figure 1): 1) free molecular flow is defined for *Kn* > 10.0; 2) transitional flow regime exists when 0.1 < *Kn* < 10.0; 3) 0.01 ≤ *Kn* ≤ 0.1 signifies the slip flow regime, and; 4) values of *Kn* below 0.01 indicate continuous flow (Tsien, 1946; Anderson, 2006; Moreno-Ibáñez et al., 2018). The first three flow regimes are important in rarefied gas dynamics with particular application to meteors (Silber et al., 2017a).

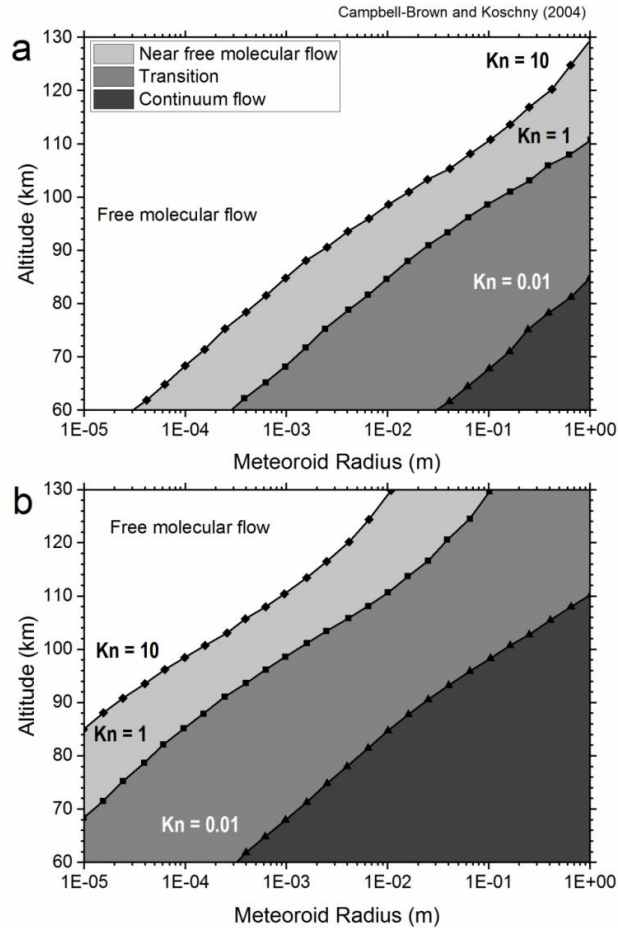

Figure 1: Flow regimes for meteoroids in the size range from $10^{-5}$ m to 1 m in diameter, assuming (a) no increase in density over the atmospheric density, and (b) intensive ablation, which increases the density by a factor of 100. The legend in (a) is applicable to both panels. Credit: Figure digitized from Campbell-Brown and Koschny, A&A, 418, 751, 2004, reproduced with permission © ESO.

It is difficult to constrain *Kn* for cm-sized meteoroids, because the altitudes where these objects generate shock waves are actually significantly higher than those where shock waves should form according to the onset of the classical continuum flow regime, based on physical meteoroid dimensions (Silber et al., 2015; Silber et al., 2017a; Moreno-Ibáñez et al., 2018). Experimental interpretation of meteor generated shock waves demonstrates that they appear much earlier than predicted by classical gas dynamics theory (Brown et al., 2007; Silber et al., 2015; Silber et al., 2017a; Moreno-Ibáñez et al., 2018). Those observations are also in line with earlier theoretical reasoning developed and specifically adopted for the theory of meteor shock waves (Bronshten, 1983 and references therein). In principle, it is well known that all bright meteors generate shock waves at altitudes where they ablate (Bronshten, 1983); the governing parameters are presented by Probstein (1961).

In comparison to typically slow and much larger re-entry vehicles, meteor shock waves form at significantly higher altitudes in contrast to altitudes predicted by the classical definition of *Kn* and the onset of continuum flow (Gnoffo, 1999). The general appearance of meteor shock waves is in the lower portion of the classically defined transitional and slip flow regime (Moreno-Ibáñez et al., 2018). This was in a way also recognized early on (e.g., Levin, 1956; Hayes and Probstein, 1959; Bronshten, 1965, 1983). The main reason for this is the formation of ablationally amplified hydrodynamic shielding (or vapor cap) which is discussed in Section 4.

For re-entry vehicles, because they are significantly larger than a typical cm-sized meteoroid and moving at comparatively much slower velocities, selection of *Kn* is straightforward (Anderson, 2006). Of course, for a typical hypersonic re-entry vehicle (see, for example, Gnoffo, 1999; Sarma, 2000; Zel'dovich and Raizer, 2002; Anderson, 2006), the onset of the continuum flow regime, which is inherently tied to the collision rate in the gas, signifies the formation of a strong shock wave front, and a flow field between the shock wave and the body - generally referred to as the shock layer. Indeed, the existence of this phenomenon, particularly in relation to ballistic and re-entry vehicles, has been known since the mid-last century, where the shock layer is relatively thin and depends on the atmospheric density, re-entry velocity and the shape of hypersonic body (Anderson, 2006). This also implies that *Kn* for a meteoroid must be evaluated differently (Bronshten, 1983; Boyd et al., 1995; Josyula et al., 2011). However, an ablating meteoroid is different from a typical hypersonic re-entry vehicle because of the existence of a vapor cloud in front of the meteoroid. The size of this vapor cloud (or cap) is proportional to the cube of the meteoroid velocity (Öpik, 1958; Bronshten, 1983; Popova et al., 2000; Boyd, 2000; Campbell-Brown and Koschny, 2004).

For meteors, an additional complication arises due to the uncertainty in the actual size of the initial vapor shielding (which basically acts as a blunt body with a sharp interface to the surrounding air, when sufficiently dense). From an analytical perspective, another challenging aspect of the meteoroid in a rarefied flow is that the continuum equations of gas dynamics fail for sufficiently large *Kn* (Boyd, 2003). Large *Kn* means that the gas does not undergo a sufficient number of collisions to maintain equilibrium velocity distribution functions (Boyd, 2003). It is commonly assumed that failure of the continuum approach occurs at Knudsen numbers around 0.01 (Boyd et al., 1995).

The solution to the general problem of evaluating object flow regimes in the rarefied atmosphere (as it applies to meteors) was recognized early on (Bronshten, 1983) and reiterated by Boyd et al. (1995) and Josyula and Burt (2011). They stated that it is better to consider local length scale and *Kn* associated with the flow field near the meteoroid instead of classically defined flow regimes. Comprehensive reviews of the hypersonic flow and discussion about associated complexity of the physico-chemical processes have been given by Gnoffo (1999), Sarma (2000) and Josyula and Burt (2011). More extensive reviews and the studies of rarefied gas dynamics and physical aspects of shock waves in non-continuum flow regimes can be found in work done by Hayes and Probstein (1959), Kogan (1969), Cercignani (2000), and Shen (2005).

Here we use the classical hypersonic dynamics terminology and consider the velocity of the hypersonic flow in terms of the Mach number, $M_\infty$ (defined as the ratio of the flow velocity to the local speed of sound) (Anderson, 2006). In respect to meteoroids, the hypersonic flow can then be referred to as the flow of atmospheric gas over the meteoroid for which $M_\infty$ values range between 35 and 270 (e.g., Boyd, 2000). As we shall see, due to very high Mach numbers and the inherent strong ablation, the treatment of meteor hypersonic flow differs significantly in physical terms from typical supersonic or hypersonic re-entry vehicles flow. The following section is concerned with the formational dynamics and implications of the meteor vapor cap (or hydrodynamic shielding).

**4. Hydrodynamic Shielding and Implications for the Shock Wave**

The hydrodynamic shielding in front of the ablating meteoroid (in some older literature defined as a vapor cap or vapor cloud, and here we use the terms interchangeably) is formed as the incident atmospheric molecules and atoms collide with the meteoroid surface, producing ejected, evaporated and reflected meteoric constituents and collisionally modified atmospheric atoms. The mean free path of reflected molecules and atoms in the vapor cloud is also a function of Mach number, local atmosphere mean free path and meteoroid-atmosphere temperature ratio, and as such, can be expressed in those terms (see Bronshten, 1983). The constituents of this cloud may have velocities up to $1.5v_{meteor}$ (Levin, 1956; Mirtov, 1960; Rajchl, 1969). As described by Mirtov (1960), there are two categories of molecular velocities: 1) "slow" velocity, due to ablated molecules escaping at thermal velocities the surface of the meteoroid (thus mainly of meteoric composition), and 2) "fast" velocity, associated with molecules of the surrounding gas leaving the surface of the meteoroid after elastic rebound, and as such exceeding the velocity of the meteoroid itself. The latter is given by $v_{particles} = v\,[1 + (1 - a)^{0.5}]$, where $a$ is the accommodation coefficient ($a = 0.75$ for iron and $a = 0.90$ for stone) (Levin, 1956; Mirtov, 1960; Rajchl, 1969). In the case when the mean free path within the hydrodynamic shielding is an order of magnitude less than the mean free path of the ambient atmosphere, the shielding effect of the cloud (Figure 2) becomes important for most meteoroids (Popova et al., 2000, 2001). The vapor cloud formation in turn initiates aerodynamic and thermal shielding, where the role of convective and radiative heat transfer during the meteoroid ablation becomes far more significant, especially following the formation of the shock wave (Öpik, 1958; Bronshten, 1983). For clarity, we note that the term meteoroid herein refers to the physical object itself, and not the object plus the vapor cloud.

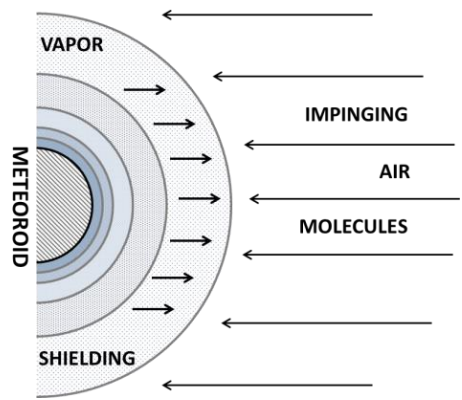

**Figure 2**. Schematic depiction of vapor shielding (after Popova et al., 2000, 2001). The contour lines and darker shading illustrate density gradients due to intensive evaporative ablation of the meteoroid and decelerated and dissociated entrained atmospheric molecules. In the regions with increasing density, higher order collisions play important role. Figure not to scale.

Let us address the question of how such phenomena affect the Knudsen number, and by extension, the formation of shock waves. The presence of hydrodynamic shielding complicates the situation in two ways: 1) the species density within the shielding is progressively increasing toward the meteoroid (Popova et al., 2000), causing the mean free path in the flow field near the meteoroid to significantly decrease (in the frame of reference of the moving meteoroid (Figure 1)); 2) The relatively large density of the shielding (which itself is a function of altitude, meteoroid velocity, composition and size) prevents the direct impacts of the local atmospheric constituents on the meteoroid, and also increases the apparent dimensions of the moving structure sometimes more than two orders of magnitude relative to the initial dimensions of the meteoroid (Boyd, 2000). Therefore, the dense vapor (Figure 3) creates near continuum flow conditions around the meteoroid (Boyd et al., 1995) and consequently alters the consideration of the local Knudsen number, $Kn_l$, in the flow field around the meteoroid (see Bronshten, 1983 for discussion). This of course takes place at altitudes where the $Kn$ associated with the local atmosphere corresponds to the transitional flow regime.

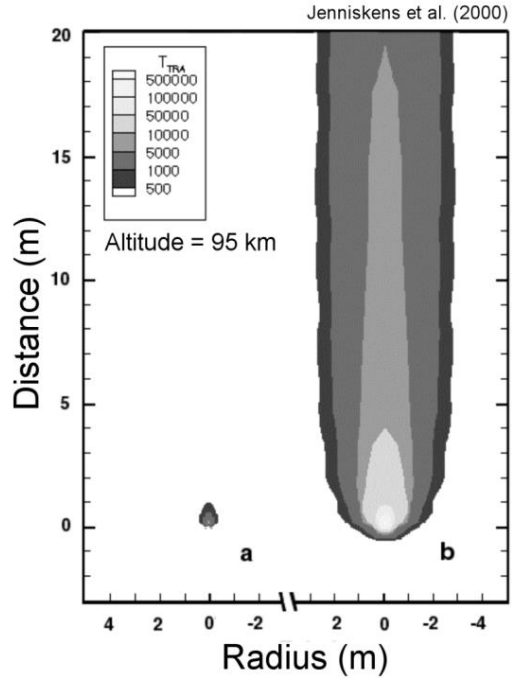

Figure 3: Translational temperature field from a rarefied flow model of a 1 cm-sized, -1 magnitude Leonid meteoroid (a) without ablation and (b) with ablation of Mg atoms. The modeling was done using the direct simulation Monte Carlo technique (see Jenniskens et al. (2000) for details). The simple ablation model follows Bronshten (1983). Credit: Jenniskens et al. (2000), Leonid Storm, reproduced with permission © Springer.

The first point discussed above enables the application of the continuum flow when the mean free path in the vapor cloud surrounding the meteoroid is such that $Kn_l$ corresponds to the aforementioned flow regime (in the reference frame of the moving object). This also allows the application of Navier-Stokes equations (Anderson, 2006) and eliminates the need for more computationally cumbersome use of the Boltzmann equation in the treatment of flow over the meteoroid. The second point is also very important because a larger area of the local atmosphere swept by the system of the moving meteoroid and the surrounding hydrodynamic shielding (Figure 3) shifts the flow regimes to higher altitudes and decreases the Knudsen number (Silber et al., 2017a and references therein; Moreno-Ibáñez et al., 2018).

However, there are large uncertainties in assessing the ablationally augmented size of the hydrodynamic shielding which effectively sweeps the region of the incident atmosphere, beyond the initial meteoroid dimensions. For example, Popova et al. (2000) estimated the size of the modeled vapor cap ahead of a centimeter-sized Leonid meteoroid and simulated a density increase in the 'vapor layer'. They obtained density values of more than two orders of magnitude higher than the density of the local region of the atmosphere. The size of the modeled vapor cloud, obtained by computational methods, is more than one order of magnitude larger than the initial meteoroid dimensions, yet relatively small in comparison with high resolution observations of the structure around a Leonid of the same size (Jenniskens et al., 2004a, Jenniskens et al., 2004b).

On the other hand, Boyd (2000) simulated a 1 cm fast meteoroid associated with the Leonid meteor shower (which is known for its high impact velocity, ~72 km/s) at an altitude of 95 km ($Kn = 4$) and showed that during its interaction with the atmosphere, the ablated vapor cloud size significantly exceeds two orders of magnitude relative to the meteoroid particle (e.g., Jenniskens et al., 2000). The result obtained by Boyd (2000) can be interpreted to agree well with the more recent study (Li et al., 2015) that presented the modeling results of a much slower flow ($M_\infty = 10$) over a sphere-cone satellite space craft at $Kn = 1.618$. However, an important observation by Popova et al. (2000, 2001) indicates that below an altitude of 114 km, the pressure of the vapor cloud in front of the meteoroid will be always greater than the ambient air pressure for all bodies with sizes 0.1 mm – 100 mm. The implication is that for most meteoroids with size ($d_m \geq 4$ mm), the (local) continuum flow condition applies at altitudes below approximately 90 km. Additionally, in the case of a Leonid, Boyd (2000) calculated that each impinging air molecule will release about 500 meteoric particles, assuming that the meteoritic composition is chondric. For a typical cometary particle, that number is up to five times greater depending on the velocity. Boyd's (2000) work agrees well with Bronshten's (1983) analytical treatment. In addition to producing very high-temperature chemically reactive flows, such a large number of evaporated meteoric atoms affect the overall dimensions of the aerodynamic shielding, and subsequently the formation of the shock layer.

The problem of the vapor cap and its role in hydrodynamic shielding, including pressures, density and energy distribution, has been further considered by Popova et al. (2000, 2001, 2003). Their simulations confirmed that below the specific altitudes associated with the onset of intensive evaporative ablation, a dense vapor cloud forms in front and around the meteoroid which screens its surface from the direct impacts of air molecules. Their results are in line with previous theoretical arguments (Levin, 1956; Lebedinets et al., 1969; Rajchl, 1969, 1972; Bronshten, 1983). In the context of their simulation results, Popova et al. (2000) also discussed the important role of UV radiation from the compressed and intensely heated air and ablated vapor in the hydrodynamic shielding layer and extending flow field. The optical thickness of the hydrodynamic shielding strongly varies as a function of the ratio of the vapor cloud specific radial mass to the mass of non-thermal air within the vapor cap. The values of optical thickness calculated by Popova et al. (2001) are between $10^{-2}$ to $10^4$. Up to 90% of the impinging air molecules end up entrapped in the meteor flow field while the rest escape during initial collisions with vapor (Popova et al., 2001). A recent modeling study by Johnston et al. (2018) examined radiative processes involved in the passage of large (1 – 100 m), low velocity (< 20 km/s) meteoroids at altitudes below 50 km. Based on their study, it can be expected that the optical thickness of ablating meteoroids at high altitudes is low. As outlined above, the main difficulty in constraining the meteor flow regime is the existence of the hydrodynamic shielding that forms in front of the meteoroid as a product of high energy collisions with the local atmospheric constituents. Initial direct collisions with the meteoroid surface causes the ejection of atoms from the lattice on the surface of the meteoroid (Öpik, 1958; Bronshten, 1983), followed by the subsequent evaporation of meteoroid surface atoms, which is the generally dominant mechanism of mass loss after the formation of hydrodynamic shielding. Collisional kinetic energy is at least an order of magnitude greater than the dissociation energy of atmospheric molecules; thus it is reasonable to assume that the impinging atmospheric gases are dissociated upon colliding with either the surface of the meteoroid or the constituents of the vapor shielding. Of course, the number of atmospheric incident molecules that are directly reflected from the meteor surface (albeit collisionally modified) is comparatively small relative to the ejected and vaporized meteoric atoms. It should be mentioned that the second and third order collisions (in addition to existing but less frequent higher order ones) take place in the dense region of hydrodynamic shielding.

Bronshten (1983) derived a conservative analytical expression that approximates the number of meteoroid molecules and atoms ejected/evaporated during each collision with an atmospheric molecule reasonably well. He first considered the collision of a singular atmospheric molecule impacting the surface of the meteoroid (at typical meteor velocities). The energy of that molecule is then written in the canonical form as: $\frac{m_a v^2}{2} = N_v Q m_m$. The LHS is the kinetic energy, where $m_a$ is the mass of the air molecule and $v$ is the velocity of the air stream over the meteoroid. On the RHS, $N_v$ is the number of vaporizing meteoric molecules/atoms, $Q$ is the latent heat of vaporization and $m_m$ is the typical mass of meteoric molecule/atom. Solving for the number of evaporated meteoric atoms, one obtains:

$$N_v = \frac{v^2}{2Q}\frac{m_a}{m_m} \qquad (8).$$

In addition to the latent heat of vaporization, the number of evaporated meteoric atoms depends on the ratio of the mass of the air molecule and the mass of the meteor atom and the meteoroid velocity. Considering that a spherical meteoroid surface is a reasonable approximation, it is clear that not all evaporated and ejected meteoric atoms end up in front of the meteoroid (Bronshten, 1983). Indeed, their location in the flow field depends on the angle at which the atom is ejected relative to the axis of the meteoroid propagation. In principle, the hydrodynamic shielding is considered effective when the mean free path of atoms and ions surrounding the front of the meteoroid along its axis of propagation is approximately an order of magnitude smaller than the radius of the meteoroid (Popova et al., 2003). Evaporation, as one of the ablational mechanisms, is the strongest in the case of strong shielding (Bronshten, 1983). In the absence of vapor shielding, the frictional force of the oncoming airflow causes spraying of the liquefied meteoroid surface layer. This is a dominant mechanism of mass loss, especially for the slower meteoroids. For the sizes ($d_m \geq 4$ mm) and velocities considered here, such ablational mechanics can be neglected for the purpose of this review, while we note that the fragmentation is an important factor of meteoroid mass loss, and much has been said about it in the literature (e.g., Artemieva et al., 1996, 2001). For the purpose of this expository review, we will only briefly revisit the topic of fragmentation in Section 8.

Zinn et al. (2004) showed that by the time the vapor cloud volume has swept up a mass of air equal to its own mass, its velocity will have decreased by a factor of two, which results in a decrease of the kinetic energy transferring internal energy to the cloud. It should be noted that the evaporated and entrained constituents of the hydrodynamic shielding cloud (which have both axial and radial components of the velocity in the expanding flow field) collisionally decelerate to thermal velocities of the ambient atmosphere in several hundred meters (Zinn et al., 2004). Rajchl (1969; 1972) also considered the vapor cloud shielding in front of the meteoroid and determined that it affects the ionization efficiency coefficient and consequently the number density of free electrons in the meteor trail. In principle, the vapor or hydrodynamic shielding by reflected, ejected or evaporated atoms and molecules is an important precursor to the appearance of shock waves (see Hayes and Probstein, 1959; Probstein, 1960, 1961; Bronshten, 1965, 1983; Silber et al., 2017a). Indeed, based on the increased density of species and augmented dimensions of the vapor cloud, it is clear why the flow regime boundaries (with respect to the meteoroid) are shifted to higher altitudes. This will subsequently affect the timing and formation of the meteor shock wave front (Jenniskens et al., 2000).

We need to note that the excitation, dissociation and ionization, along with radiative effects, take place within the hydrodynamic shielding (because of initial high energy collisions). The impact of such processes on shock formation and structure will be discussed shortly. However, the magnitude of such contributions strongly depends on meteoroid size, composition, velocity and of course, the Knudsen number. Additionally, an increase in velocity also shifts the flow regime of meteoroids upward. If the velocity increases from 20 km/s to 60 km/s, the flow regime boundaries shift upward between 8 - 12 km in altitude (e.g., see Bronshten, 1983 for discussion). The hydrodynamic shielding affects the drag coefficient, heat transfer and ablation coefficient. A detailed discussion about the dependency of those parameters on the vapor cap is given by Bronshten (1983). For example, the value of the drag coefficient is proportional to the Knudsen number and it decreases with decreasing _Kn_ (as the atmosphere becomes denser). This happens because the formation of hydrodynamic shielding reduces or eliminates direct collisions of the incident atmospheric molecules and atoms with the meteoroid (Öpik, 1958; Larina, 1975; Perepukhov, 1967; Bronshten, 1983).

The heat transfer coefficient is dependent on the meteoroid velocity and type of flow regime, as well as on the corresponding mechanism of heat transfer that dominates in the particular flow regime (e.g., behind the shock wave front, shock layer radiation is a significant component of meteoroid heating) (Baldwin et al., 1971; Revelle, 1979; Stulov, 1997). Behind the dense hydrodynamic shielding, the heat transfer consists of both convection and radiative heating (e.g., Strack, 1962; Nerem, 1965). After formation of the hydrodynamic shielding (or vapor cloud), as the altitude decreases, the heat transfer coefficient also decreases (Popova, 2005). This is well illustrated for the case of a 1 cm Leonid in the transitional flow at 90 km altitude. For example, the heat transfer coefficient for a Leonid at this altitude is a factor of 5 lower than that if the Leonid was in the free molecular regime (Popova, 2005). The ablation coefficient ($\sigma_a = \Lambda/2QC_D$) also depends on the meteoroid velocity (Jacchia, 1958; Verniani, 1967; Bronshten, 1983; Popova, 2005 and references within), atmospheric density and the rate of evaporation. The estimated value of ablation coefficient ranges from 0.001 s$^2$/km$^2$ to 0.21 s$^2$/km$^2$ (e.g., see Revelle, 1979; Revelle et al., 2001; Gritsevich, 2009; Bouquet et al., 2014).

The heat flux received by a unit surface of the meteoroid increases with exponentially increasing atmospheric density (and the density of the hydrodynamic shielding) until the formation of the strong shielding. In the denser atmosphere below 100 km altitude, the magnitude of heat flux (which of course depends on meteoroid size and velocity) becomes much greater than the heat released by the meteoroid through radiation (Bronshten, 1983). The intense evaporation behind either a shock wave or hydrodynamic shielding becomes the dominant mechanism of ablation (for simplification, we do not consider removal of liquid droplets, fragmentation, etc.). Details of such mechanisms are discussed by Bronshten (1983).

Another important parameter to consider in discussion of the flow regimes is the Reynolds number ($Re$) which characterizes the ratio of inertial forces to viscous forces in a fluid flow. The Reynolds number can be written as:

$$Re = \frac{v \rho_a d_m}{\mu} \qquad (9),$$

where $v$, $d_m$ and $\rho_a$ are the meteoroid velocity, characteristic dimensions of meteoroid and density of the ambient atmosphere, respectively. The symbol $\mu$ represents dynamic viscosity. We note that in the case of strong ablation, the problem at hand would be more complicated (e.g., see Figure 3), and as a result, $d_m$ in Eq. (9) would be replaced with the vapor cloud diameter. This expression for $Re$ is slightly different than that used in atmospheric research, where the Reynolds number represents the ratio of the turbulent viscosity ($K_V$) and the kinematic viscosity ($v$), i.e., $Re = K_V/v$. With regard to the meteoroid Reynolds number, if $Re \leq 1$, then viscous forces play a significant role and must be included in analytical or numerical treatment of the flow fields and subsequent formation of the shock layer. On the other hand, if $Re \geq 1$, the flow tends to be inviscid.

In terms of Reynolds number, $Kn_v$ within the vapor cap can be written as follows (Bronshten, 1983):

$$Kn_v \approx \frac{1}{Re} \frac{\bar{v}_v}{c} \qquad (10),$$

where $\bar{v}_v$ is the mean velocity of the reflected/evaporated atoms and molecules from the meteoroid surface and generally considered within the frame of reference of the vapor cloud. The speed of sound ($c$) is expressed as $c = \left(\gamma \frac{kT}{m_a}\right)^{0.5}$. Here, $\gamma$ is the specific heat ratio, $k$ is the Boltzmann constant and $T$ is the temperature. The mass of the atmospheric molecule is denoted by $m_a$. The mean velocity of the evaporated molecules and atoms can be reasonably approximated as a function of the meteoroid surface temperature ($T_s$) and written as $\bar{v}_v = \left(\frac{8kT_s}{\pi m_m}\right)^{0.5}$. However, as we mentioned earlier, the velocity of reflected/ejected/evaporated atoms may be greater than the velocity of the meteoroid, in the case of specular reflection/ejection and when considered from the frame of reference of the surrounding atmosphere. It should be noted that for a 1 cm meteoroid moving at 30 - 70 km/s, *Re* is in the range of $10^2$ or greater below 90 km (increasing with decreasing altitude) (Bronshten, 1983). While the assumption of inviscid flow is generally valid for the hydrodynamic shielding, this consideration changes for the shock wave (essentially for the case of the boundary layer, to be discussed further in the text) (Anderson, 2006).

## 5. Transition from Hydrodynamic Shielding to Shock Wave

The transition from the compressed vapor cloud to the shock front occurs with the meteoroid's descent to the lower altitudes (in classically defined lower transitional and slip-flow regimes). In the simplest sense, this takes place when the compressed vapor, due to the effects of the drag forces from the denser atmosphere, collapses towards the meteoroid and when the pressure and density inside the collapsed region are much higher than that of the ambient air (when a local Knudsen number corresponds to the continuum flow regime). In a strict sense however, the transition from the vapor cloud to the shock front takes place when the magnitude of the jump in pressure, density and temperature inside of the hydrodynamic shielding is large relative to the ambient air, thereby creating a discontinuity in those parameters at the shock front (Boyd et al., 1995). At this point, a rapid translational and radial expansion of the high temperature flow field behind the shock envelope around the meteoroid can be treated as a hydrodynamic flow in a vacuum (e.g., Kornegay, 1965; Masoud et al., 1969; Kustova et al., 2011). Consequently, the flow within the meteor flow field (which is encompassed by the shock envelope) can be appropriately described by the Navier-Stokes equations for compressible flow (Hayes and Probstein, 1959).

In principle, the formation of the discontinuity that corresponds to the shock wave must satisfy the Rankine-Hugoniot equations (Zel'dovich and Raizer, 2002; Sachdev, 2004) (which relate the upstream and downstream values of density, bulk velocity, and temperature in an ideal compressible fluid), leading to the formation of the compressed shock layer and viscous boundary region close to the body (Probstein, 1961; Bronshten, 1983). The Rankine-Hugoniot equations (see Section 7.1.1) were derived independently by William John Macquorn Rankine in 1869 and Pierre-Henri Hugoniot in 1887 (Ben-Dor et al., 2000). These relations apply just as well to the shock waves as to blast waves, because they express the conditions at the shock front, which is treated as a discontinuity (Zel'dovich and Raizer, 2002; Needham, 2010). While the Rankine-Hugoniot equations are generalized conditions across the shock wave, for a viscous, high temperature reacting gas, considerations are more complicated because of factors such as variable heat capacity ratio (discussed further in the text) (Anderson, 2006).

Some researchers such as Probstein (1961) reasoned theoretically that the onset of the shock wave is a gradual process, where the shock front, viscous shock layer, boundary layer and other shock features of the meteor shock wave (See Section 6.2) do not form instantaneously. In this respect, Probstein (1961) presented analytical arguments that the formation of the shock wave begins when the compressed layer has a thickness of the three local mean paths within the initial vapor cap before the transition to the shock. However, this may be only a reasonable approximation as the vapor cap exhibits strong density, temperature and pressure gradients (Popova et al., 2000, 2001). Whether the onset of shock formation is "instantaneous" or gradual still remains unresolved in the case of meteors. In principle, Bronshten (1983) pointed out that the shock exists and has clearly defined typical features such as a viscous and boundary layer when the apparent shock thickness is about 90 times smaller than the characteristic dimensions of a meteoroid.

## 6. Meteor Generated Shock Waves

A simple way to conceptualize a meteor shock wave is to think of it in terms of a snow plough analogy (Bershader, 1960; Masoud et al., 1969). The gas dynamics analogy to a meteor generated shock wave is the shock wave generated by a hypersonic blunt body. The generic term used for a shock wave associated with a hypersonic blunt body is the bow shock and it is also the term sometimes used in reference to meteor shock waves. While in principle this is not an incorrect terminology, further clarifications need to be made for meteor generated shock waves. This is primarily because of major differences between a typical hypersonic blunt body and a meteoroid that has a much higher Mach number of propagation, strong ablation and very low Mach cone angle $\Phi = \sin^{-1}(1/M_\infty)$, which justifies the approximation of the initial paraboloid shock envelope as a cylinder. Another major difference between "regular" hypersonic blunt bodies and meteor shock waves is that the shock wave associated with the latter is analogous to the shock from a cylindrical line source (Lin, 1954; Bennett, 1958; Sakurai, 1964; Jones et al., 1968; Plooster, 1968, 1970).

The meteor shock wave is essentially a paraboloid surface that can be divided, for pedantic purposes, into two distinctive shock regions as follows:

i. Along the axis of meteor propagation and closely in front of the meteoroid is a roughly hemispherical shock region analogous to fore body bow shock (sometimes referred to as the ballistic shock) in hypersonic blunt bodies, that forms first when the hydrodynamic shielding collapses and densities, velocities and temperatures can be treated as discontinuities across the shock front. Here, instead of fore body bow shock, we refer to it as the shock in front of the meteoroid (essentially a hemispheric surface of high temperature, pressure and flow velocity discontinuity, in front of the meteoroid). This is also a meteor shock region with maximum dissociation, excitation and ionization effects. Let us recall that the strength of the shock wave is directly proportional to the pressure behind the shock front, primarily because the pressure is a mechanical variable (Anderson, 2006). The pressure behind the shock front is directly proportional to the free stream velocity and for a high speed meteoric flow where $v_2 \ll v_1$, and $p_2 \gg p_1$ and can be approximated as $p_2 \approx \rho_1 v_1^2$. The subscripts 1 and 2 denote the parameters associated with the free stream and the shock layer, respectively.

ii. The bow shock wave behind the meteoroid (a continuous transition from the shock wave in front of the meteoroid) (also see Section 6.2), which is essentially a surface of a paraboloid meteor shock envelope in the far field, and is herein referred to as a cylindrical shock wave (Sakurai, 1964, 1965). It should be reiterated that the cylindrical shock wave is essentially a continuous extension of the shock envelope that begins with the shock in front of the meteoroid discussed above. Correspondingly, an additional clarification needs to be made here. The primary cylindrical shock wave (basically the bow shock wave) propagates radially outward and it is trailed by a strong and ablationally amplified recompression shock wave (Silber et al., 2017a).

For clarity of this exposition, we do not distinguish here between the bow and recompression shock waves. However, the distinction is made in Silber et al. (2017a), and illustrated in Section 6.2. It must be emphasized that in the case of meteors, both types of shock waves are ablationally amplified and as they coalesce rapidly, they cannot be individually distinguished beyond the boundaries of the initial radius of the meteor train (in terms of spatial coordinates) and especially outside of the immediate region of maximum energy deposition with the characteristic radius $R_0$ (to be defined shortly) (ReVelle, 1974, 1976; Silber et al., 2015; Silber et al., 2019). Here, we refer to both the initial meteor bow shock wave and a recompression driven shock wave as simply a cylindrical shock wave. Indeed, meteor cylindrical shock waves in the strong shock regime are the primary topic of this review. Cylindrical shock waves attenuate rapidly and transition to the weak shock regime within approximately ten characteristic radii ($R_0$) (ReVelle, 1974, 1976; Silber et al., 2015). Since this topic is well covered in literature, only a limited discussion on weak shock is given in the infrasound section (Section 10.2).

*6.1 Shock Definition*

Before further discussion about the nature and dynamics of meteor shock waves continues, we shall establish the definition of shock waves based on the classical shock wave literature. For the purpose of completeness of this expository review, we present several different definitions of the shock waves.

First, we need to note that any object moving at velocities higher than the local speed of sound of the medium creates a disturbance which is called a shock wave. This is true when the density of the medium and characteristic dimension of an object satisfy the conditions for a continuum flow regime. The term wave, however, implies a time dependent phenomenon. In shock waves, the time dependent phenomenon is the region of the flow immediately behind the shock front, where thermodynamic parameters change rapidly (Hurle, 1967). Therefore, in conceptual terms, one can define the shock wave as the region of rapid change of thermodynamic parameters of the gas parcel flow immediately ahead of the cylindrical shock front. Specifically, we are interested in the change from their initial 'equilibrium' to their 'final' values directly behind the shock front. However, in reality the matter is far more complex (Cercignani, 2000; Brun, 2009, 2012).

In a more technical sense, a shock wave can be described as a discontinuous surface that connects supersonic flow with subsonic flow (Niu, 2009). After the atmospheric molecule impacts and passes through the shock wave front, its flow velocity is then dramatically reduced and can no longer be considered as part of the hypersonic flow (see Zel'dovich and Raizer, 2002; Anderson, 2006). Therefore, an irreversible adiabatic compression with accompanying sharp decrease of velocity and a jump in density behind the shock front cause almost instantaneous increase in pressure, temperature and entropy. In principle, the flow behind the shock front is in non-equilibrium, and irreversible processes occur inside the shock layer (Niu, 2009).

Finally, following the most formal definition (Ben-Dor et al., 2000), shock waves can be characterized as *mechanical waves of finite amplitudes that arise when matter is subjected to a rapid compression*. The characteristic properties of shock waves can be distinguished by: *(i) a pressure-dependent, supersonic velocity of propagation; (ii) the formation of a steep wave front with abrupt change of all thermodynamic quantities; (iii) for nonplanar shock waves, a strong decrease of the propagation velocity with increasing distance from the center of origin; and (iv) nonlinear superposition (reflection and interaction) properties*. In the strict mathematical sense however, shock waves cannot be treated as simple discontinuities in rarefied gasses (Cercignani, 2000). Zel'dovich and Raizer (2002) discussed this and stated that from a mathematical perspective, a shock wave discontinuity can be regarded as the limiting case of much larger but finite gradients in flow variables across the shock layer whose thickness tends to zero.

An analysis of shock waves, their dynamics, nature and types is given, for example, by Hayes and Probstein (1959), Liberman et al. (1986), Ben-Dor et al. (2000), Cercignani (2000), Zel'dovich and Raizer (2002), Sachdev (2004), Anderson (2006), Krehl (2009), Brun (2012), among others. Additional works covering relevant and shock related aspects of high temperature and rarefied gas dynamics are presented by Kogan (1969), Shen (2005), Brun (2009) and Bose (2014). However, our emphasis returns now to aspects of the meteor generated shock waves, which are the primary focus of this paper.

At this point it is important to emphasize that due to the initial high translational temperatures and strong amplification by ablation (vaporization), the shock waves produced by a meteoroid are significantly stronger than the comparative shock wave in the case of no ablation and/or from a re-entry vehicle.

For typical meteoroid velocities, a massive amount of flow kinetic energy in a hypersonic free stream is converted to internal energy of the gas across the strong shock in front of the meteoroid, hence producing very high temperatures in the shock layer region. High temperatures ($T \geq 11,000$ K) in the front shock layer (Anderson, 2006) (including the sonic and boundary region which will be defined shortly), along with densities and pressures that exceed that of the ambient air by several orders of magnitude (Bronshten, 1983), are instrumental in approximating meteor cylindrical shock waves as explosive line sources (e.g., Lin, 1954; Bennett, 1958; Sakurai, 1964; Jones et al., 1968; Plooster, 1968; Tsikulin, 1970).

This approximation is valid as a large amount of energy is released per unit length in a finite cylindrical volume (Lin, 1954; Plooster, 1968; Tsikulin, 1970; ReVelle, 1974, 1976; Steiner et al., 1994). The radius of a cylindrical region with the maximum meteor energy deposition per unit length is referred to as the characteristic or blast radius ($R_0$) and can be expressed as: $R_0 = (E_0/p_0)^{0.5}$, where $E_0$ is the energy deposited per unit path length (which in the case of a meteoroid is the same as the total aerodynamic drag per unit length) and $p_0$ is the ambient pressure (e.g., Tsikulin, 1970; ReVelle, 1974, 1976; Silber et al., 2015; Silber and Brown, 2019). There are slight variations to the expression for $R_0$ (Sakurai, 1964; Jones et al., 1968; Few, 1969; Plooster, 1968; Tsikulin, 1970; also see Infrasound, Section 10.2), resulting in a difference by a factor of ~3.5 (ReVelle, 1974; Silber, 2014; Silber and Brown, 2019). The term characteristic radius is used only in reference to strong shock waves, when the energy release ($E_0$) is sufficiently large so that the internal energy of the ambient atmosphere is negligible (Lin, 1954; Hutchens, 1995). However, practical real time detections of meteor generated shock waves at the characteristic altitudes of formation in the upper atmosphere (for $d_m \geq 4$ mm), have not been possible up to this point because of the rapid spatial and temporal attenuation of the shock waves in the rarefied atmosphere, complicated further by the presence of luminous phenomena.

*6.2 Meteor Shock Wave Morphology*

Figure 4 represents a schematic of a meteoroid shock envelope and flow fields. In the next few paragraphs, bracketed numbers will refer to Figure 4 unless otherwise stated. The paraboloid surface of shock wave envelope (2) that surrounds the initial meteoroid (1) flow field is the extension of the shock front (3) ahead of the meteoroid.

The detached and relatively thin shock front (Taniguchi et al., 2014), characterized by a large gradient in temperature, pressure and velocity, occurs where the viscous and conductive transport phenomena dominate. Because the shock front is generally very thin, the fluid element (i.e., a molecule) can only experience a limited number of high energy collisions and because of the short time scale, the effects of those collisions (such as vibrational equilibration, dissociation, ionization) take place behind the shock front. Thus, we can say that the flow through the shock front is chemically frozen. However, that is not always the case and it gets progressively more complicated with the presence of strong radiative phenomena, which modify the ambient air near the meteoroid and along the axis of its propagation.

In a general analysis of the shock waves, the value of $\frac{\rho_2}{\rho_1}$ plays an important role in the determination of the shock detachment distance $\delta$ for the hypersonic body. It is generally acceptable to express the ratio of the shock detachment distance and the specific radius of the body $r_m$, in terms of density ratios (see Anderson, 2006), and it can be written as:

$$\frac{\delta}{r_m} = \frac{\frac{\rho_1}{\rho_2}}{1+\sqrt{2(\frac{\rho_1}{\rho_2})}} \qquad (11).$$

In the limit of high velocities, the shock detachment distance becomes very small because $\frac{\rho_1}{\rho_2} \ll 1$. Here, we can indirectly see the effects of chemistry (i.e., dissociation) on the shock detachment distance. As dissociation and consequently the number species density increases, this leads to the decrease of the shock detachment distance (Bronshten, 1965; Zel'dovich and Raizer, 2002; Anderson, 2006).

When the shock front ahead of the meteoroid transitions into the much weaker and radially expanding cylindrical shock wave, then the streamlines of particles entrained behind the original meteor shock front become part of the cylindrical shock wave. Correspondingly, the Mach number of the streamlines in the shock flow field at the edge of the boundary layer (5) has a strong influence on the stability of the laminar boundary layer (Beckwith, 1975; Anderson, 2006) at the transition region where flow becomes turbulent in the meteor flow field. The region behind the detached shock front consisting of sonic (4), viscous boundary (5) and stagnation region (6) layers, is fundamentally a thin region with very strong gradients of pressure, kinetic temperature and density resulting from continued adiabatic compression of the vaporized meteoric atoms and impinging atmospheric gas.

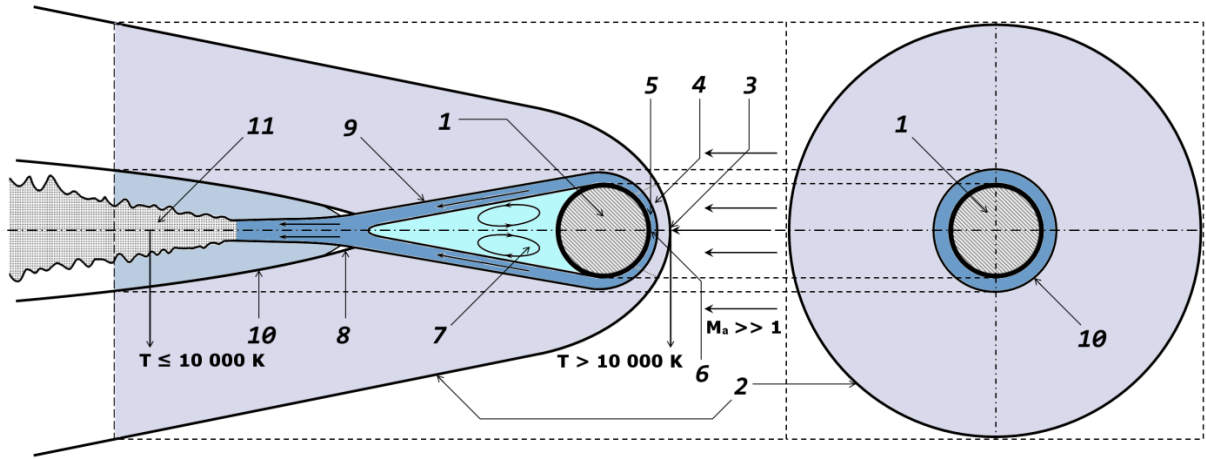

**Figure 4**: Schematics of the meteor shock wave(s), flow fields and near wake. The meteoroid is considered as a blunt body (with the spherical shape) propagating at hypersonic velocity from left to right. The left side represents the side view, while the right side represents the head-on view (after Hayes and Probstein (1959); Lees et al. (1962)). The definitions and explanations are provided in the text. (1) Meteoroid; (2) Cylindrical (bow) shock wave envelope; (3) The shock front; (4) Sonic region; (5) Boundary layer; (6) Stagnation point; (7) Turbulent region (in some older literature, this is referred as the dead water region); (8) The neck and recompression region; (9) The 'free' shear layer; (10) The recompression region (a source of recompression and ablationally amplified cylindrical shock wave) shock wave front; (11) The region of turbulent "plasma" flow and adiabatic expansion. The diagram is only for the illustrative purpose and is not to scale.

At this time it is useful to recall that shock properties are going to be governed by chemistry, pressure and temperature. The choice of specific heat ratios (*which depend on chemistry, temperature, pressure*) affects the geometry of the initial shock front region (Bronshten, 1965; Zel'dovich and Raizer, 2002; Anderson, 2006) and shock layer thickness (Figure 5). Additionally, meteoroid shock wave thickness depends on dissociation rates of atmospheric molecules and evaporation rates (Zel'dovich and Raizer, 2002). The assumption of $\gamma = 1.4$ is appropriate for approximations; however, it is not valid in rigorous analytical and numerical treatment of the shock wave, as can be seen in Figure 5 (Steiner and Gretler, 1994). The effects of dissociation, electronic excitation and ionization must be considered when choosing appropriate specific heats ratio for the analysis of meteor shock waves and associated flow fields. Notably, the gas specific heat ratio affects the thickness of the boundary layer.

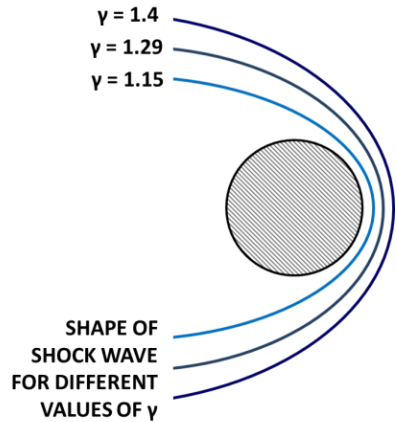

**Figure 5.** The dependence of shock wave shape and thickness on the various values of ratio of specific heats (after Bronshten, 1965). Direction of meteoroid propagation is left to right. Note that the diagram is not to scale.

The flow field between the shock front and the meteoroid is defined as the shock layer, and for hypersonic speeds this shock layer can be quite thin. The shock layer in front of the meteoroid is characterized by an almost instantaneous increase in temperature, density and entropy (Anderson, 2006) and consists of the sonic region (4), where the flow velocity is decelerated to subsonic values and of the viscous boundary layer (5). The boundary layer is the region adjacent to a surface where the effects of molecular friction (viscosity), thermal conduction and radiative heating are dominant. The flow outside of the boundary layer can be treated as inviscid. The viscous boundary layer envelopes the meteoroid and is a region where non-equilibrium processes occur (Zel'dovich and Raizer, 2002; Anderson, 2006). In particular, the thickness of this layer depends on the mean free path, chemical reactions, and the ratio of the specific heats.

The shock layer becomes infinitesimally thin and infinitely dense in the limit where $M_\infty \to \infty$ and $\gamma \to 0$. The specific heat ratio plays a key role in the study of shock waves properties. To illustrate this, consider (in the simple case of an ideal gas), for example, the limiting density or a volume ratio across the shock wave is a direct function of $\gamma$, as can be seen from $\frac{\rho_2}{\rho_1} = \frac{V_1}{V_2} = \frac{\gamma+1}{\gamma-1}$, where $\rho_1$ and $V_1$ are initial density and volume of the upstream flow and $\rho_2$ and $V_2$ correspond to the values of density and volume of the compressed region behind the shock wave (see Zel'dovich and Raizer, 2002). However, at high pressures and temperatures, the specific heats and their ratio are no longer constant. As we have seen from an earlier discussion, this is due to the presence of intensive dissociation, ionization, radiative and other nonequilibrium processes (Zel'dovich and Raizer, 2002).

In shock waves, viscosity and heat conduction are a dissipative process and are functions of the molecular structure of a 'fluid' (Zel'dovich and Raizer, 2002). Such waves are generally associated with the boundary layer and appear only in the case of large gradients in the flow variables (e.g., velocity, density, temperature). It should be added that the thickness of the shock layer is greater at higher altitudes (e.g., for illustrative examples see Li et al., 2015). Moreover, the shock layers may be optically thin, characterized by the absence of re-absorption of radiation emitted from other parts of the gas. On the other hand, in an optically thick shock layer re-absorption of radiation is significant and radiation must be considered as a separate source of energy. Correspondingly, the modeling of optically thick gases is extremely difficult, since, due to the calculation of the radiation at each point, the computation load theoretically expands exponentially as the number of points considered increases (e.g., Sarma, 2000).

The highest translational temperatures ($T \gg 11{,}000$ K) within the shock layer are found in a comparatively small stagnation region (6) which is localized within the boundary layer (and also part of the sonic region). However, the heating in the stagnation region varies inversely with the square root of the characteristic radius of the meteoroid. Hence, bigger meteoroids will not reach temperatures much greater than cm sized ones (because of much greater radius of the blunt region). This was also experimentally verified (Koppenwallner, 1984). Zinn and Drummond (2005) showed numerically that high temperature regions persist in the coma at significant distances (Figure 6).

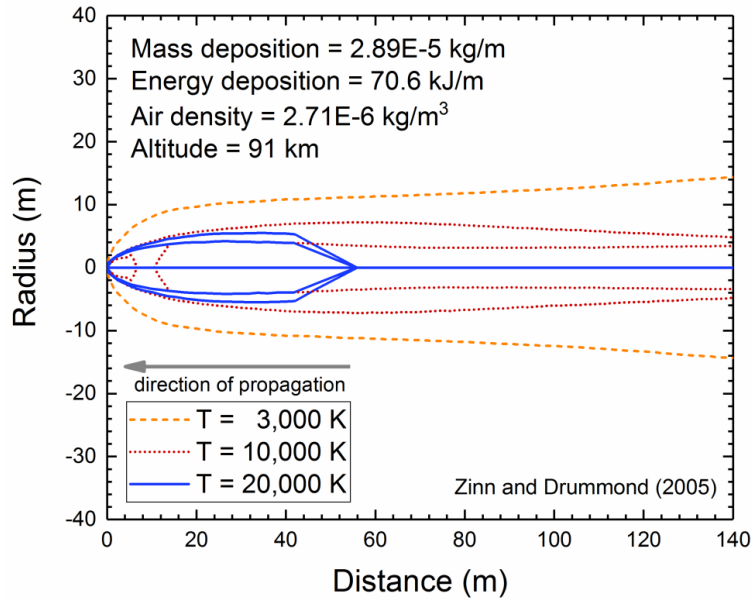

**Figure 6:** Plot showing modeled contours in the wake of a 526 g Leonid meteor with an entry angle of 42° (Zinn and Drummond, 2005). Credit: digitized Figure 5 from Zinn and Drummond (2005), JGR, 116, A04306, reproduced with permission © John Wiley and Sons.

High meteor temperatures have been observationally confirmed in meteor coma and imply the presence of local thermodynamic equilibrium (LTE) (e.g., Berezhnoy and Borovička, 2010). Harvey (1973) observed the temperatures of 20,000 K, while Ceplecha (1973) observed spectral lines of atomic oxygen corresponding to temperatures of about 14,000 K in meteor wakes. Borovička (1994) concluded that meteor wake temperatures of 10,000 K are relatively common, and represent the temperature of only a small fraction of the ablated meteor flow field volume. However, Borovička (1994) stated that such high temperatures are due to meteor shock waves. This is consistent with conclusions made from the observation of the Perseid fireball spectrum in which a high temperature component was determined to originate as a result of meteor shock (Borovička et al., 2006). Jenniskens et al. (2004a) obtained the spectrum emitted from the meteor trails corresponding to temperatures of about 4,400 K and demonstrated that the thermalization is significantly slower than previously theoretically considered. They concluded that the variation of meteor plasma emission temperatures for meteoroids in the range of masses between $10^{-5}$ g and $10^6$ g is only up to several hundred Kelvin.

In comparison to initially enormous kinetic temperatures exceeding tens of thousands of K (Boyd, 2000; Silber et al., 2017a), these observations can be attributed to energy loss as a result of radiative emissions and intense dissociative, exciting and ionizing collisions behind the shock. These collisions lead to internal energy exchange between translational and internal degrees of freedom and result in excitation, dissociation, ionization and radiation. In the compressed shock layer, the excess internal energy in turn facilitates various non-equilibrium and hyperthermal chemical reactions (Berezhnoy and Borovička, 2010; Panesi et al., 2011; Brun, 2012) further down in the reactive high temperature meteoroid flow fields (Zel'dovich and Raizer, 2002; Anderson, 2006; Brun, 2009).

We shall digress for a moment here for the purpose of completeness. Let us define $\tau_f$ as a characteristic time for a fluid element to travel the distance of the flow field, approximated by the ratio of the characteristic length of the flow field and upstream velocity of the gas before the shock front, and $\tau_c$ as the characteristic time for the chemical reactions. The "equilibrium flow" (see Anderson, 2006) is characterized by $\tau_f \ll \tau_c$. A non-equilibrium flow is defined in all other cases (Anderson, 2006).

Inadequate choice of specific heats ratio (e.g., $\gamma = 1.4$) can be the source of large errors in calculating the temperatures behind shock wave. This is best illustrated if we compare temperatures in the shock layer of the Apollo lunar re-entry module. The calculated translational temperature behind the shock wave ($M_\infty = 32.5$; $h = 53$ km) is 58,128 K. The actual temperature is about 11,600 K (Anderson, 2006). Therefore, it is important to use chemically reacting real gas in calculations as it prevents large over-prediction in the calculated temperature. That is because the directed kinetic energy of the flow, when converted across the shock wave, is shared across all molecular modes of energy, and utilized in zero-point energy of the products of chemical reaction. Thus, we want to emphasize again the importance of strong effects that dissociation, ionization, radiation and hyperthermal chemistry processes have on the temperature reduction behind the meteor shock wave.

As noted earlier, if the shock-layer temperature is high enough, the fluid elements in the flow will emit and absorb radiation. This causes the meteor flow field to become non-adiabatic. Moreover, strong radiation increases pressure in the shock layer by influencing the dissociation rates. For general hypersonic velocities greater than 7 km/s (e.g., Kemp, 1959; Keck et al., 1959; Kivel, 1961; Camm et al., 1961; Strack, 1962; Nerem, 1965; Reis, 1967; Anderson, 1969; Levin et al., 1993; Erdman et al., 1993) the energy is converted to radiation in the stagnation region (Bronshten 1983). At meteor velocities, radiative heat transfer is more efficient in comparison to convective stagnation-point heat transfer (Anderson, 1967, 1969; Sutton, 1985). However, radiative cooling also has an important role in the reduction of high kinetic temperatures behind the shock front. Anderson (1967) presents a major survey of radiative shock layer effects associated with hypersonic flight dynamics.

At meteor velocities, a large component of the radiation from the radiating volume behind the shock front is in the UV spectrum and is responsible for preheating and modifying the atmospheric region ahead of the meteor axis of propagation, where high energy photon fluxes from the shock layer are absorbed by the ambient air (Cook et al., 1960). The mean free path of UV radiation between 100 – 120 km altitude ranges from about 300 to 3,000 m, respectively (Zel'dovich and Raizer, 2002). The radiation from the stagnation region is also responsible for heating the meteoroid, as well as for additional modification of the meteor flow field. Strong radiation (including UV) from the meteor shock region (the strongest radiating volume is within the stagnation region) is responsible for the ionization and dissociation of the local atmosphere swept by some solid angle ahead of the meteoroid axis of travel (Cook and Hawkins, 1960; Jenniskens et al., 2004a). This mechanism effectively creates more dense local atmosphere in front of the meteoroid axis of propagation and effectively decreases *Kn* (some discussion is given in ReVelle (2001)). However, in the specific case of meteors, while the observational spectral evidence shows strong UV emissions from meteors (Borovička et al., 1998; Carbary et al., 2003), especially the region behind the meteor shock, the quantitative assessment of the flux of high energy photons and the role of the ambient atmosphere modification ahead of the meteoroid is still poorly understood (Jenniskens et al., 2004a). One of the reasons for this is that the precise values of optical thickness of the meteor shock layer remain relatively uncertain at the moment.

Strong radiative processes from the shock layer (Figure 4) that contribute significantly to the heating of the meteoroid body and vaporization process behind the shock layer have been extensively considered by several authors (Nemchinov et al., 1963; Stulov, 1972; Gershbein, 1974; Zel'dovich and Raizer, 2002). Additionally, radiation also affects the value of $\gamma$ (Zel'dovich and Raizer, 2002) and consequently influences the shock layer thickness and mean free path by dissociation of molecules that initially might have passed through the shock front undissociated.

We now again refer to Figure 4. Note the bracketed numbers in the following paragraphs again refer to the numbers in Figure 4. Behind the meteoroid, the boundary layer transitions into a 'free' shear layer (9), which exhibits strong gradients in velocities between the outer and inner flows. In both regions, the ablated meteor vapor, mixed with the entrained collisionally modified atmospheric gases, is driven toward the recompression region (or the neck) (8). The maximum compression of the high temperature non-equilibrium aft flow takes place in the neck region (8). Further down in the flow (behind the meteoroid), this region is also the source of the strong recompression shock (10) for strongly ablated bodies where the flow velocity has both radial and axial components. The recirculation zone (7) is driven by the pressure gradients, and the flow circulates back toward the aft end of the body and outward toward the shear layer separation point (Gnoffo, 1999). The flow field of the ablated and evaporated material encapsulating the meteoroid and extending into the near wake can be described by the Navier-Stokes equations for continuity, momentum and energy. While equations for continuity and momentum are purely mechanical and not affected by chemical processes, the energy equation needs to take into consideration viscous, thermal and chemical effects in the hypersonic gas flow (Anderson, 2006).

Further back in the wake (11), the ablated meteor vapor and plasma experience initially strong turbulence (Lees and Hromas, 1961, 1962) during the adiabatic expansion into a more dynamically stable volume with radius $r_0$ (discussed earlier) that is in pressure, but not in temperature, equilibrium with the ambient atmosphere. Thermodynamic equilibrium cannot be achieved in the meteor flow field (including the near wake). However, only a local thermodynamic equilibrium (which is highly altitude dependent) may be achieved (Berezhnoy and Borovička, 2010). In rarefied flow conditions, i.e., for small Reynolds numbers, the shock waves and viscous layers are thick (e.g., see numerical model by Li et al., 2015). Consequently, under such conditions a classical gas dynamics approach for the analysis of flows and shock waves may be challenging to implement.

*6.3 Pressure Behind the Shock Wave – Initial Approximation*

The ratio of pressures behind the shock wave and in the upstream of the ambient atmosphere is the most important and singular parameter that determines the strength and subsequent effects of the meteor generated shock wave. Let us see how we can approximate it initially.

As mentioned earlier in this section, a simple way to conceptualize a meteor shock wave is in terms of a snow plough analogy (Bershader, 1960; Masoud et al., 1969). Another way to visualize a meteor shock wave is to think of a meteor flow field, driven radially by a hypersonically expanding piston (e.g., see discussion in Lin, 1954; Bennett, 1958; Sakurai, 1964; Jones et al., 1968; Anderson, 2006). A strong cylindrical shock wave propagates radially from the point of an 'instantaneous' energy release. This again illustrates why meteor shock waves are often referred to as cylindrical blast waves and why they are analogous to the blast wave theory as developed by Sedov (1946) and Taylor (1950). Of course, we know that the situation is significantly more complicated, as one needs to account for the strong ablation and extremely high flow field temperatures and pressures (i.e., Zinn et al., 2004; Zinn and Drummond, 2005). The following discussion is helpful in addressing the more detailed treatment of meteor shock waves later in the text.

In the case when the pressure ahead of shock wave can be neglected, asymptotic formulas for velocity, density, and pressure near the center of the explosion (in the strong shock regime) can be obtained from self-similarity principles developed by Sedov (1946) and Taylor (1950). For example, the pressure resulting from cylindrical blast waves (Sedov, 1946; Lin, 1954; Sakurai, 1964) can be expressed as a function of deposited energy (essentially drag resistance per unit length):

$$p = K\rho_0 \left(\frac{E}{\rho_0}\right)^{0.5} t^{-1} \qquad (12),$$

where $K = \dfrac{\gamma^{[\frac{2(\gamma-1)}{2-\gamma}]}}{2^{[\frac{4-\gamma}{2-\gamma}]}}$. From the equivalence principle (Sedov, 1946; Taylor, 1950; Anderson, 2006) we can define time $t$ as $t = \dfrac{x}{V_0}$. The above expression (Eq. (12)) for pressure is in good agreement with experimental results (Lukasiewicz, 1962; Sakurai, 1953, 1954). The pressure distribution (ratio) in the case of shock waves produced by a hypersonic blunt cylinder, derived from the first approximation (see Anderson, 2006 for discussion) can be expressed as:

$$\frac{p}{p_0} = 0.067 M_\infty^2 \frac{\sqrt{C_D}}{(x/d)} \qquad (13),$$

where the subscript 0 denotes the conditions ahead of the blast wave, $x$ is the distance traveled by the blunt body in some time interval, $d$ is the diameter of the blunt object and $M_\infty$ and $C_D$ are the Mach number of the flow and drag coefficient of the blunt body, respectively. It should be noted that because of the factor ($x/d$) this expression allows an approximation of the pressure in the meteor wake, just before the equilibration with the ambient atmosphere (which coincides with the formation of the meteor trail with initial radius $r_0$). For a blunt-nosed body, the pressure distribution $p/p_0$ is directly proportional to $M_\infty^2$ and $C_D^{1/2}$, and inversely proportional to $x/d$ (Anderson, 2006).

As we shall see shortly, these first order analytical and empirical approximations (Anderson, 2006) used in hypersonic dynamics of re-entry bodies provide reasonable approximations for ablating meteors, however they do not account for the effects of high temperature and meteoric ablation. We should keep in mind though that these are only approximations and for actual meteor flow fields, their values might be understated because of large uncertainty associated with hydrodynamic shielding and the initial shock envelope effect on $C_D$, the variable specific heat ratio and strong ablation and non-equilibrium flows which impact species density in the flow field behind the shock. While, for example, the pressure behind the shock wave in front of the meteoroid is mainly a function of drag force, the pressure (which characterizes the strength of the shock wave) in the flow field behind the meteoroid that drives the cylindrical shock is a function of ablation rate, chemistry and temperature.

### *6.4 Meteor Kinetic Energy Transferred to the Ablated Vapor*

In terms of an analytical treatment of the meteor cylindrical shock wave, it can be approximated using a geometric argument (see Anderson, 2006). A better approach, at least in the region of the strong shock, is to use similarity principles (Sedov, 1946; Taylor, 1950) and to evaluate the shock wave based on the energy released per unit length (or cylindrical line source). In the case of meteors, because of the strong ablation and strong energy deposition along the axis of propagation by ablated vapor (Zinn et al., 2004), the second approach is better.

Approximately 99% of energy deposited by the meteoroid per unit length and in a "confined" cylindrical volume comes from collisional stoppage of ablated compressed vapor. The latter expands behind the shock envelope and has both a radial and axial component of velocity (Zinn et al., 2004). This is in line with earlier estimates of the meteor kinetic energy partition and conversion (Romig, 1964). Zinn et al. (2004) pointed out correctly that the "drag energy" of the solid meteoroid transfer to the impinging atmospheric gas is negligible in comparison. This can be also understood in the context of the loss and partition of the meteoroid kinetic energy which can be written as the loss of meteoroid kinetic energy:

$$\frac{dE}{dt} = \frac{d}{dt}\left(\frac{mv^2}{2}\right) = mv\frac{dv}{dt} + \frac{v^2}{2}\frac{dm}{dt} \qquad (14).$$

Here, the first term on the LHS represents the energy lost per unit of time, and *m* and *v* are the meteoroid mass and velocity, respectively (e.g., Romig, 1964; Gritsevich and Koschny, 2011). Dividing both sides by the velocity (*v*) (Bronshten, 1983) the energy deposition per unit path length can be obtained:

$$\frac{dE}{dl} = m\frac{dv}{dt} + \frac{v}{2}\frac{dm}{dt} \qquad (15).$$

The first term on the right in Eq. (15) is the energy used to form the bow shock wave, assuming no ablation. The second term then is the energy partitioned to the ablation and lost to the ablated vapor per unit length and accounts for the $\frac{dm}{dt}$. After some manipulation, this can be expressed as:

$$\frac{v}{2}\frac{dm}{dt} = \frac{dE_{vapor}}{dl} = \frac{A}{2}\rho_a v^4 \qquad (16),$$

where *A* is a function of the drag coefficient, the cross-sectional area of the meteoroid, the heat transfer coefficient and the heat of vaporization, and is also different for meteoroids with different properties (see Zinn et al., 2004 for derivation). This was originally considered by Dobrovol'skii (1952), subsequently verified by other researchers (Bronshten, 1983) and implemented in numerical modeling (Zinn and Drummond, 2005). Bronshten (1983) showed that the second term on the RHS of Eq. (15) is utilized to describe the formation of the comparatively stronger ablationally amplified meteor shock wave. Taking the ratio of the RHS terms in Eq. (15) viz., $\frac{v}{2}\frac{dm}{dt} / m\frac{dv}{dt}$, the second term may be up to two orders of magnitude larger than the first term (Dobrovol'skii, 1952; Bronshten, 1983). Of course, this ratio depends on the meteoroid velocity, composition and the rate of ablation.

We should emphasize that the energy required to completely vaporize the meteoroid is negligible compared to its initial kinetic energy. As can be seen from the discussion above, it is the rapid deceleration and expansion of the high temperature ablated vapor which drives and amplifies the strength of meteor cylindrical shock waves (Zinn et al., 2004). Let us again revisit the pressure behind the shock wave envelope of ablating meteor.

### *6.5 The Pressure Ratio Revisited*

The pressure ratios of the ablated, vaporized meteoroid and plasma, mixed with dissociated atmospheric gases in the flow field (*p*) to that of the ambient atmosphere ($p_0$) for a meteoroid with 1 cm radius, are in the range $10^2 < p/p_0 < 10^4$ (Bronshten, 1983). This is particularly true for events with velocities exceeding 30 km/s, where much more energy is transferred to the flow field vapor and plasma behind the shock front, than is spent on the ablation process. It is the dispersion of this ablated and pressurized "vapor" in the front of the meteoroid that amplifies the shock wave (Dobrovol'skii, 1952; Bronshten, 1983; Zinn et al., 2004). The experimental evaluation of the effects of pressure ratios (in the similar range as above) on the velocity and strength of the shock waves was performed by Kornegay (1965). Typical energies of cm-sized meteoroids, assumed to be released instantaneously per unit length along the axis of propagation, may readily exceed several kJ/m (Zinn et al., 2004; Silber et al., 2015).

The strength and therefore speed of a meteor produced shock wave principally depends on the pressure difference between the pressure behind the initial shock envelope and the ambient pressure (e.g., Hurle, 1967; Bronshten, 1983; Zel'dovich and Raizer, 2002; Anderson, 2006). In practice, *p/p₀* can be relatively easy to compute based on initially known meteoroid parameters (Bronshten, 1965; Tsikulin, 1970; Bronshten, 1983). For one-cm chondritic meteoroids moving at average meteor velocities, at the stage of maximum ablation, *p/p₀* term readily reaches $10^4$ (Bronshten, 1983). Several factors contribute to the increase in pressure jump behind the initial meteor shock envelope and affect the strength of the meteor cylindrical shock wave. Those are the meteoroid size, velocity and the initial flow translational temperatures, rate of ablation and classically determined Knudsen number value. Initial high temperatures behind the meteor shock front that readily exceed 11,000 K (Anderson, 2006) contribute, together with a high number density of ablated/evaporated meteoric atoms and entrapped impinging shock dissociated atmospheric molecules, to the pressure increase that drives the meteor shock wave. We emphasize again that pressure can easily reach four orders of magnitude higher than that of the ambient pressure (Bronshten, 1983).

The cylindrical shock wave velocity (or Mach number) and subsequently the shock strength can be easily obtained from the classically derived expression for the pressure behind the shock front which is generally evaluated using the Hugoniot relationship. This relates the vapor pressure behind the shock ($p$) and the ambient pressure ($p_0$) to the product of shock Mach number ($M_{sw}$) and the ratio of specific heats ($\gamma$) (e.g., Lin, 1954; Jones et al., 1968; Tsikulin, 1970):

$$\frac{p}{p_0} = \frac{2\gamma}{\gamma+1} M_{sw}^2 \qquad (17).$$

This relationship can be used in the region of the strong shock wave where $p \gg p_0$ (Lin, 1954; Jones et al., 1968). Because pressure is a "mechanical" variable (see Anderson, 2006), the assumed value of $\gamma$, corresponding to an ideal gas, will not significantly alter the value of the ratio in the expression above. However, for temperature, as we shall see, the situation is quite different (Anderson, 2006).

The empirically derived relations for the density, pressure and temperature ratios (see Zel'dovich and Raizer, 2002) that describe well the experimental observations in the region of the strong shock wave (Jones et al., 1968; Plooster, 1970 and references therein) are written as:

$$\frac{\rho'}{\rho_0} = \frac{6}{1 + 5M_{sw}^{-2}} \qquad (18a),$$

$$\frac{P'}{P_0} = \frac{7}{6} M_{sw}^2 - \frac{1}{6} \qquad (18b),$$

$$\frac{T'}{T_0} = \frac{1}{36}(7 - M_{sw}^{-2})(M_{sw}^2 + 5) \qquad (18c).$$

Here, following the notation of Zel'dovich and Raizer (2002). $\rho_0$, $p_0$, $T_0$ are the density, pressure and temperature ahead of the shock wave, respectively; and $\rho'$, $p'$, $T'$ are the values just behind the shock front. These equations (18a-c), alone or in combination, may be used to obtain the unknown parameters behind the strong cylindrical shock wave (for details see Zel'dovich and Raizer, 2002; Hurle, 1967). However, while Eqs. (18a,b) are reasonably accurate in the strong shock regime, caution must be exercised with the closed analytical form of Eq. (18c), as the temperature ratio might be considerably overestimated (Anderson, 2006).

While the in-depth review of meteor shock waves concludes at this juncture, we shall continue our discussion on analytical and numerical treatment of meteor hypersonic flow (Section 7), meteoroid fragmentation (Section 8), shock waves in the context of larger objects that produce airbursts and pose impact hazard (Section 9), as well as radar and infrasound detections of meteor generated shock waves (Section 10). These sections will provide the reader with a broader picture and understanding of shock wave related phenomena.

## 7. Analytical and Numerical Treatment of Meteor Hypersonic Flow

The main challenge in analytically and computationally analyzing meteor flow fields and shock waves in a rarefied atmosphere (where $Kn$ is high and shifts to the transitional flow regime) is that the continuum equations are also invalid in regions of very low density. Thermal non equilibrium is another aspect of rarefied flows. The validity of the continuum equations is inherently tied to the collision rate in the gas (Josyula and Burt, 2011). At high altitudes, the rarefied atmosphere does not behave like a continuum medium, and when a relatively small number of particles occupy the flow field of interest, the continuum hypothesis is obviously invalidated, as large regions of the flow field may contain no particles at all. This problem and analysis of shock waves in a rarefied environment is discussed by Cercignani (2000) and Lago et al. (2012).

Thus, the question is: when does a continuum approach fail in the treatment of shock fronts and how does that affect the shock waves in a rarefied flow? Boyd (2003) discussed the parameters that predict the continuum breakdown, and the efficiency and validity of those criteria under specific conditions. The existing breakdown parameters that are based on the flow field gradients are unable to predict a breakdown in the shock front because the Navier-Stokes equations can only model very thin shocks. However, the parameter based on the local Knudsen number that describes the flow field, rather than the classically defined *Kn*, is more reliable to predict continuum breakdown at the body surface and in the downstream at the edge of the shock wave (where the Navier-Stokes equations are still valid).

In the frame of reference of the meteor shock wave in the rarefied atmosphere, much denser meteor ablated vapor and plasma expand in the rarefied ambient atmospheric gas. Considering initial large density pressure and temperature gradients in the flow field in the meteoroid near-wake, such radial expansion of the meteor vapor and plasma behind the shock wave can be approximated as a continuum flow into vacuum, until the point where the flow is rarefied and the continuum breaks down. It must be emphasized again that in the case of ablating meteoroids, a strong cylindrical shock wave is driven by "dispersion" of high temperature ablated vapor and plasma (Bronshten, 1983; Zinn et al., 2004). Thus, for a strongly ablating meteoroid following the formation of the shock wave, the flow regime within the flow field bound by the initial shock envelope is a continuum; however, in the wake of the meteoroid, the flow field expands and becomes rarefied. In this region, the continuum equations are not valid and the Boltzmann equation (Boltzmann, 1872) should be used. The Boltzmann equation here is the governing equation in the microscopic description of a dilute gas flow, and is a nonlinear integro-differential equation for a probability distribution function that statistically describes the state of the particles as a function of time (Boyd et al., 1995; Cercignani, 2000).

The main disadvantage of the Boltzmann equation is that it can offer analytical solution only in a very limited set of simple problems. For example, the linearized Boltzmann equation is only possible in some cases where the disturbance is small enough such that the velocity distribution function is perturbed only slightly from the equilibrium or Maxwellian form (Prasanth et al., 2006). However, such cases have no real applicability in hypervelocity rarefied flow conditions (e.g., re-entry vehicles or meteoroids). The conditions arising from shock waves produced by hypervelocity meteoroids are inevitably highly non-linear with large perturbations, thus necessitating a numerical approach (see Section 7.2).

Continuum equations will hold until the meteor flow field has sufficiently expanded in the wake (before the formation of the dynamically stable volume of plasma with initial radius $r_0$). Within those spatial and temporal boundaries, the propagation of the strong cylindrical shock wave can be treated as a blast into vacuum. After that the continuum equations break down.

## 7.1 Analytical Approach

### 7.1.1 Euler's Equations of Fluid Dynamics

The general equations that govern inviscid high temperature flows are called Euler equations of fluid dynamics (Anderson, 2003, 2006, 2005). The Euler equations are strictly based upon the Maxwellian velocity distribution, which is the prevailing distribution when the gas is in equilibrium. Furthermore, as they include no viscous terms, they are only valid at very low Knudsen numbers, or when the collision rate becomes quite large. When this occurs, the gas is able to redistribute its energy to accommodate varying conditions almost instantly, and the flow is practically in equilibrium everywhere. Thus, in the case of meteors, Euler's equations can only be used in the continuum flow regime (or used in a transitional flow regime as an approximation, assuming local thermodynamic equilibrium in the flow of the strongly ablating meteor flow field that satisfies the continuum conditions).

Euler's equations, in the case of an inviscid flow of an ideal gas, give a series of closed–form algebraic relations for the pressure and temperature ratios. Derivations of the basic continuity (total mass conservation), momentum (basically a second Newton's law) and energy (basically a total enthalpy in this case) Euler equations for an inviscid flow field are given in Anderson (2003, 2005). Their solution depends on the boundary and initial conditions for the specific flow that is considered. Here, we are just going to write those following Anderson (2006) notations:

Continuity:

$$\frac{d\rho}{dt} + \nabla \cdot (\rho \mathbf{V}) = 0 \qquad (19a),$$

Momentum:

$$\rho \frac{D\mathbf{V}}{Dt} = -\nabla p \qquad (19b),$$

Energy:

$$\rho \frac{Dh_0}{Dt} = \frac{\partial p}{\partial t} \qquad (19c).$$

Here, $\rho$ and $p$ are density and pressure, and $\mathbf{V}$ is velocity of the flow field (in vector form). Total enthalpy per unit mass ($h_0$) can be expressed as $h_0 = h + v^2/2$. The energy equation for the inviscid flow that involves entropy (describes adiabatic flow) is not valid in this case, as entropy increases irreversibly in inviscid non-equilibrium high temperature flows. Use of the entropy term is justified in so called "equilibrium" flows which are just approximations and not especially realistic in the case of meteor flow fields. That is the reason that the entropy term has been replaced by total enthalpy in this form of meteor flow field description. Because the continuity equation is 'mechanical' in nature, it remains the same for all considerations, including chemically reacting, non-equilibrium and viscous flows. However, the energy Eq. (19c) does not hold for non-equilibrium inviscid flows, which will be briefly discussed soon.

Under the conditions of the strong shock wave, let us assume that the local thermodynamic equilibrium (defined earlier) is valid behind the shock front. Then across the normal shock wave surface (as in the case of the shock front for larger meteoroids) we can obtain relations for the conservation of mass (or continuity), momentum and energy, also known as the Rankine-Hugoniot equations. These are easily derived from the Euler equations for steady, one-dimensional flow, by integrating between points in front of and behind shock. Then we can write them as follows:

Continuity:

$$\rho_1 v_1 = \rho_2 v_2 \qquad (20a),$$

Momentum:

$$p_1 + \rho_1 v_1^2 = p_2 + \rho_2 v_2^2 \qquad (20b),$$

Energy:

$$h_1 + \frac{v_1^2}{2} = h_2 + \frac{v_2^2}{2} \qquad (20c).$$

The subscript 1 corresponds to the undisturbed flow ahead of the shock wave and the subscript 2 denotes the quantities in the compressed shock layer. These equations hold well for both reacting and non-reacting gases. In this case density and temperature behind the shock wave are the function of pressure and energy behind the shock front, or $\rho_2 = \rho(p_2, h_2)$ and $T_2 = T(p_2, h_2)$.

For an ideal gas, for the pressure and temperature ratio behind and in front of shock wave (as a function of Mach number) we can obtain a series of closed-form relations. On the other hand, for the case of non-ideal gas in a reactive flow, with strongly excited vibrational energy, the equations above must be solved numerically. Therefore, in the case of reacting flows (e.g., Brun 2009) the pressure, density and energy ratios behind and in front of a shock wave, are functions of velocity, pressure and temperature in the upstream flow: $\frac{p_2}{p_1} = f_1(v_1, p_1, T_1)$, $\frac{\rho_2}{\rho_1} = f_2(v_1, p_1, T_1)$, and $\frac{h_2}{h_1} = f_3(v_1, p_1, T_1)$.

The basic inviscid flow equations (19a-c) are modified in the presence of non-equilibrium chemistry or radiative heating. In the presence of non-equilibrium chemistry, Eq. (19c) has an additional term, $\frac{dw_i}{dt}$ (e.g., see Anderson, 2003, 2006, 2005), which corresponds to the local rate of change of the chemical species (*i*) as a result of chemical reactions inside the considered volume element; this needs to be considered. This is represented by molar mass ($M_i$) of species (*i*) multiplied by its chemical rate of change ($\frac{dX_i}{dt}$): $\frac{dw_i}{dt} = M_i \frac{dX_i}{dt}$. Therefore, the continuity equation (for specific species in the flow) is now $\frac{d\rho_i}{dt} + \nabla \cdot (\rho_i V) = \frac{dw_i}{dt}$; however, the general continuity equation stays the same (e.g., see Anderson, 2003, 2006, 2005). Here *V* is the vectoral component of the flow velocity. The energy equation (for example, in the case of volume heating by radiation absorption) includes an additional term $\frac{dq}{dt}$; $\rho \frac{Dh_0}{Dt} = \frac{\partial p}{\partial t} + \frac{dq}{dt}$.

It should be also noted, counter-intuitively, that increasing the pressure behind the shock wave may impede the dissociation and ionization efficiency, while the temperature ratio $\frac{T_2}{T_1}$ is higher at higher pressures. To illustrate the problem, in Figure 7 we show the temperature ratio at different atmospheric pressures and in the range of lower velocities up to 14 km/s (Anderson, 2006). However, the meteor flow field, following the formation of the shock wave, cannot simply be treated as an inviscid flow (e.g., Euler's equations of fluid dynamics). The picture is complicated if we consider the presence of high temperatures, as well as viscous and chemically reactive non-equilibrium flow (depending on the regions in the shock wave layer and the general location in the flow field). To address that, one can use the Navier-Stokes equations for the viscous flow (e.g., flow in the boundary layer).

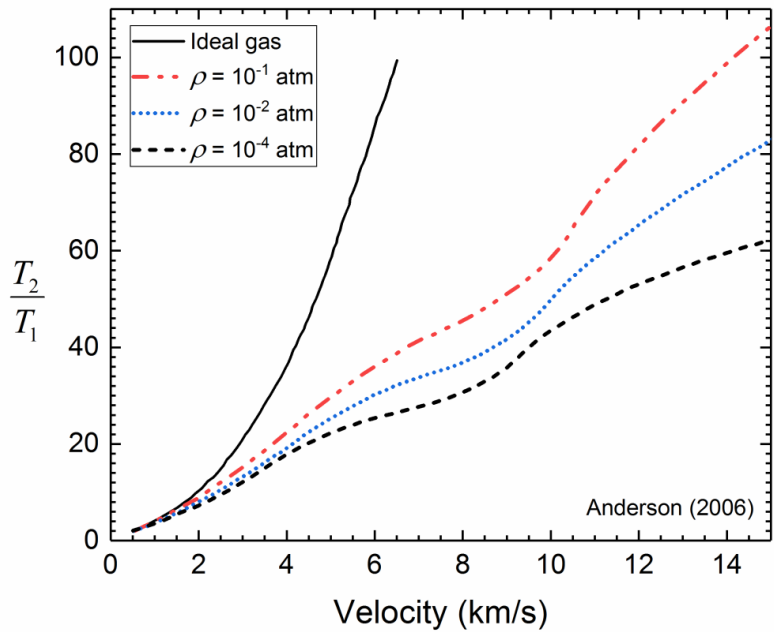

**Figure 7:** Influence of pressure on the normal shock temperature in equilibrium air. Credit: digitized Figure 14.3 from Anderson (2006), reproduced with permission © John Anderson, Jr.

At this point of discussion it is worth noting that while the Mach number is a practical tool in the evaluation of most equilibrium hypersonic flows, in the case of meteor flow fields that have high-temperature and chemically reactive flows in general, the Mach number is not a particularly useful quantity. A better approach is to treat the flow of a chemically reacting gas in terms of "primitive" variables and to use the upstream and downstream velocity, temperature, and pressure. For an assumption of equilibrium gas, the Mach number is still defined in terms of the ratio of stream velocity and the local speed of sound, and it can be used for reasonable approximations.

### 7.1.2 Navier-Stokes Equations

The governing set of equations for a compressible viscous flow are the Navier-Stokes equations. The Navier-Stokes equations have an advantage over Euler's equations because they exhibit an extended Knudsen number validity (Josyula and Burt, 2011). While the breakdown is still subject to some debate (Boyd et al., 1995, 2003), it is generally considered that this occurs when Knudsen number is in the region of 0.1 (Bird, 1994). Beyond that, the Navier-Stokes equations fail in the meteor near wake where the ablation amplified flow field starts to rarefy.

As noted, the Navier-Stokes equations are the viscous form of the Euler equations, and a common approach is to use these equations to extract similarity parameters (Taylor, 1950; Sedov, 1946). They are applicable in the boundary layer within the shock, which is characterized by frictional and conductive energy dissipation. However, these equations do not have analytical solutions. That is resolved by defining a set of boundary conditions for the viscous boundary layer. The concept of the boundary layer was first developed by Ludwig Prandtl in 1904 (Anderson, 2006). The flow outside the boundary layer is treated as inviscid.

Recall that the basic flow evolution associated with the meteor cylindrical shock wave can be described by a set of Euler's equations (e.g., see Lin, 1954; Steiner and Gretler, 1994; Hutchens, 1995), which are essentially the limiting form of the general viscous flow equations in the limit of infinite Reynolds number (Anderson, 2006). Another key difference in the Navier-Stokes equations is the presence of the shear stress tensors, associated with velocity gradients. A full derivation of the Navier-Stokes equations which define the continuity, momentum and energy is given in Anderson (2003), and a detailed discussion is in Anderson (2006). It should be noted that in the general equation of motion for a fluid flow, viscous effects do not influence the basic principle of mass conservation (Anderson, 2006). Hence, as we stated earlier, the continuity equation remains the same as the Euler equation in the case of inviscid flow.

Viscous interactions behind the shock front become particularly important at lower altitudes of meteor ablation, where we can differentiate:

1) Pressure interaction, caused by exceptionally thick boundary layers on surfaces under some hypersonic conditions; and
2) Shock wave/boundary-layer interaction, caused by the impingement of a strong shock wave on a boundary layer (Anderson, 2006).

The thickness of boundary layer grows as the square of the free stream Mach number, and therefore hypersonic boundary layers can be orders of magnitude thicker than low-speed boundary layers at the same Reynolds number. It is well known that, because of the large-scale turbulent motion and turbulent diffusion, energy is transmitted more readily in turbulent boundary layers than in laminar.

For the general case of the chemically reactive, high temperature viscous flow, the governing set of the Navier-Stokes equations can be written:

Continuity:
$$\frac{d\rho}{dt} + \nabla \cdot (\rho \boldsymbol{V}) = 0 \qquad (21a),$$

x-Momentum:

$$\rho \frac{Dv}{Dt} = -\frac{\partial p}{\partial x} + \frac{\partial \tau_{xx}}{\partial x} + \frac{\partial \tau_{yx}}{\partial y} + \frac{\partial \tau_{zx}}{\partial z} \quad (21b),$$

y-Momentum:

$$\rho \frac{Du}{Dt} = -\frac{\partial p}{\partial y} + \frac{\partial \tau_{xy}}{\partial x} + \frac{\partial \tau_{yy}}{\partial y} + \frac{\partial \tau_{zy}}{\partial z} \quad (21c),$$

z-Momentum:

$$\rho \frac{Dw}{Dt} = -\frac{\partial p}{\partial z} + \frac{\partial \tau_{xz}}{\partial x} + \frac{\partial \tau_{yz}}{\partial y} + \frac{\partial \tau_{zz}}{\partial z} \quad (21d).$$

The energy equation for hypersonic nonequilibrium and viscous high temperature flows is:

$$\rho \frac{D(e+V^2/2)}{Dt} = -\nabla \cdot \boldsymbol{q} - \nabla \cdot p\boldsymbol{V} + \frac{\partial(u\tau_{xx})}{\partial x} + \frac{\partial(u\tau_{yx})}{\partial y} + \frac{\partial(v\tau_{zx})}{\partial z} + \frac{\partial(v\tau_{xy})}{\partial x} + \frac{\partial(v\tau_{yy})}{\partial y} + \frac{\partial(v\tau_{zy})}{\partial z} + \frac{\partial(w\tau_{xz})}{\partial x} + \frac{\partial(w\tau_{yz})}{\partial y} + \frac{\partial(w\tau_{zz})}{\partial z}$$

(21e). The full derivation of this expression is given by Anderson (2003).

Here, the term $e + V^2/2$ is the total energy per unit mass, where $e$ denotes total internal energy and contains the heats of formation. The heat-flux is denoted by $\nabla \cdot \boldsymbol{q}$, which includes radiative heat $\boldsymbol{q_R}$ (Steiner and Gretler, 1994) (emitted or absorbed by the element volume). The heat flux vector is expressed as:

$$\boldsymbol{q} = -k\nabla T + \sum_i \rho_i \boldsymbol{U}_i h_i + \boldsymbol{q_R} \quad (22),$$

where $k$ is thermal conductivity and the subscript $i$ denotes the sum over different species. The term $U_i$ is the velocity of the $i^{th}$ species in the flow field, and it is generally different from other species flow field velocity. We note that each real species also contribute differently to the total specific internal energy in the equation of state (Steiner and Gretler, 1994).

The term $\boldsymbol{V}$ represents the velocity of all species, and terms $u$, $v$ and $w$ are components of velocity. The viscous terms (shear and normal stress terms) are in the form $\tau_{xy}$ and $\tau_{xx}$ (in similar fashion for other terms). The shear stress can be expressed as: $\tau_{xy} = \tau_{yx} = \mu(\frac{\partial v}{\partial x} + \frac{\partial u}{\partial y})$ where $\mu$ is viscosity coefficient. The expressions for normal stress can be written as: $\tau_{xx} = \lambda(\nabla \cdot \boldsymbol{V}) + 2\mu \frac{\partial u}{\partial x}$ where $\lambda = 2/3\mu$ and is defined as the bulk viscosity coefficient.

Following the exposition by Anderson (2006), we can say that in terms of the material derivative, Equation (21e) shows that the change in total energy of a fluid element moving along a streamline takes place because of: 1) thermal conduction across the surfaces of the fluid element; 2) transport of energy by diffusion into (or out of) the fluid element across its surfaces; 3) radiative energy emitted or absorbed by the element; 4) rate of work done by pressure forces exerted on the surfaces of the element; and 5) rate of work done by shear and normal stresses exerted on the surfaces. A more comprehensive discussion about the equation (including individual terms) (above) is given in Anderson (2006) and Anderson (2003, 2005). The Navier-Stokes continuity equation remains the same, as seen previously, because it is 'mechanical' in nature.

In the case of a blunt body (e.g., meteoroid), the flow conditions at the outer edge of the viscous boundary layer correspond to the local thermodynamic equilibrium. For a smaller class of shock-producing meteoroids, the local thermodynamic equilibrium is only established further in the downstream flow (Berezhnoy and Borovička, 2010). Correspondingly, the boundary layer can have regions of LTE, non-equilibrium and frozen flows (Anderson, 2006).

Finally, for the case of viscous and chemically reacting flow (with the presence of LTE) we can say that: $p = p(e, \rho)$ and $T = T(e, \rho)$. Both the Navier-Stokes and Euler's equations for the conservation of mass, momentum and energy can be written for cylindrical coordinates to accommodate an analysis of the cylindrical meteor shock waves. In the case of Euler equations, they are easily rewritten to accommodate cylindrical coordinates in unsteady one-dimensional flow (Lin, 1954; Steiner and Gretler, 1994; Hutchens, 1995), where only the space-coordinate (*r*) and time (*t*) are independent variables. The equations for the strong cylindrical shock wave produced by the meteor are written using the similarity assumptions developed by Sedov (1946) and Taylor (1950). Their solutions have been subsequently shown to be reasonably close to experimentally obtained results (Bennett, 1958; Jones et al., 1968) for the strong shock regime (i.e., $\frac{p_2-p_1}{p_1} > 30$ ) as defined by Plooster (1970). Steiner and Gretler (1994) and Hutchens (1995) analyzed strong cylindrical blasts in cases where the source mass is not negligible and obtained improved solutions for cylindrical shock waves.

## *7.2 Numerical Approach*

A comprehensive discussion on numerical modeling of hypersonic flow is well beyond the scope of this paper, considering that this topic is extensive, well researched, and well represented in literature (e.g., see Sarma, 2000; Bertin et al., 2003; Longo et al., 2007; Gnoffo et al., 2010; McNamara et al., 2011; Boyd, 2014; Li et al., 2015; and references therein). Nevertheless, in this section we present a brief outline of numerical approaches relevant to meteors.

It is important to note, however, that in re-entry problem, typical velocities are 7 – 12 km/s, much lower than that of incoming meteoroids (~11 – 72.5 km/s). Coupled with complexities surrounding the process of ablation, the modeling of meteoroid entry flow is a very challenging problem. Despite huge strides in improvement of computational power and sophisticated numerical models to-date, no model exists that can accurately describe ablation and hypervelocity flow of meteoroids at high altitudes and at all meteoroid entry velocities. This remains one of the most critical and underexplored areas of meteor shock wave research.

Boyd (2000) modeled the flow field around a 1 cm diameter Leonid meteoroid moving at 72 km/s at 95 km altitude (*Kn* = 4, $M_\infty$ = 270). His results were roughly consistent with spectroscopic observations of the 1998 Leonid meteor shower and indicate that material properties assumed for the meteoroid play a notable role. Boyd (2000) also discussed the challenges of modeling high speed and high altitude meteoroids, and outlined complexities involved in the problem. A study by Zinn et al. (2004) considered the interaction of a fast ~500 g Leonid with the atmosphere at 91 km altitude by numerically modeling the rates of ablation and deceleration, energy deposition and terminal altitudes. They reported the effects of the shock wave, radiative expansion and chemical reactions, and their results are consistent with direct observational data. However, no other study since had embarked into the domain of modeling meteoroid flow fields at such altitudes and velocities.

A choice of the numerical approach will depend on the flow regime considerations. While the Navier-Stokes equations are appropriate for describing the continuum flow, they fall short when describing the transitional and free molecular flows (e.g., Cercignani, 2000; Scanlon et al., 2015; Li et al., 2015). In particular, the breakdown of continuum becomes noticeable at approximately *Kn* ~ 0.1 (Oran et al., 1998), when it becomes difficult to maintain the velocity distribution when slightly perturbed from the Maxwellian distribution (e.g., Wang et al., 2003). This is because the particle nature of matter in the transitional regime (0.01 < *Kn* < 10) becomes extremely important and has to be explicitly accounted for. For example, once the average distance between particles is comparable to the characteristic length scale of the body, the Navier-Stokes equations can no longer produce adequate approximations to the physics of gas dynamics (e.g., Prasanth and Kakkassery, 2006). In particular, the linear transport terms for mass, viscosity, diffusion, and thermal conductivity in the partial differential equations are no longer valid (Prasanth and Kakkassery, 2006). For such problems it is necessary to implement the Boltzmann equation of kinetic energy, which can be used to consider the molecular transport phenomena across all Mach numbers and Knudsen numbers of rarefied gases, and as such it is considered the best approach to describe the hypersonic rarefied gas flows (e.g., Cercignani, 2000; Li et al., 2015). The most common application of the Boltzmann equation is in describing the conditions relevant to the problem of re-entry vehicles in the upper atmosphere, where hypersonic conditions in the rarefied flow produce ionizing reactions, and subsequently a thin plasma sheath which can block communications (e.g., Boyd, 2007a, 2014; Boyd et al., 2016). Other applications are discussed in (Prasanth and Kakkassery, 2006).

The direct simulation Monte Carlo (DSMC) method, first introduced by Bird (1963), is the most commonly used particle-simulation method for the Boltzmann's equation (e.g., Cercignani, 2000; Prasanth and Kakkassery, 2006; Boyd, 2007a; Scanlon et al., 2015; Li et al., 2015). DSMC is a particle method, where a large number of real gas molecules are represented by one particle. Details on the DSMC setup and modeling approach are given by Prasanth and Kakkassery (2006). The hypervelocity flow conditions (e.g., re-entry vehicles, meteoroids) also require chemical nonequilibrium models (e.g., see Boyd, 2007b; Boyd and Josyula, 2016; Hao et al., 2016). Observational data from re-entry vehicles, such as Reentry F (Carter et al., 1971), FIRE (Flight Investigation of Reentry Environment) II (Hash et al., 2007 and references therein), and RAM-C II (RAM is the Radio Attenuation Measurement) test vehicle (Jones et al., 1972), offer valuable information for validation and improvement of numerical models and are still being used today. For example, using the flight data from the RAM-C II vehicle and FIRE II capsule, Hao et al. (2016) tested two different 11-species chemical models (Gupta, 1990; Park, 1990) to examine the sensitivities of each and to find how well they describe the real flight conditions.

There are still computational challenges surrounding the DSMC method. For example, Boyd (2007b) discussed challenges in representing a plasma sheath surrounding the RAM-C II flight experiment using a DSMC code for hypersonic ionized flow conditions for a reacting gas. Despite modern improvements in computer hardware, the DSMC method can be computationally expensive, depending on the nature of the problem investigated. Moreover, the DSMC method has branched out into numerous code variants, discussion of which is beyond the scope of this paper (for further discussion, see Prasanth and Kakkassery, 2006; Josyula and Burt, 2011; Li et al., 2015). A review of physico-chemical modeling in hypersonic flow simulations is given by Sarma (2000).

As described earlier in this section, the Navier-Stokes equations are appropriate for the continuum flow ($Kn < 0.01$). The Computational Fluid Dynamics (CFD) approach is suitable for describing complex flow fields, including shock-shock interaction, shock-boundary interaction and flow separation (e.g., Gnoffo et al., 2010). An example of the pressure and temperature flow field around a meteoroid 10 cm in diameter moving at 35 km/s at altitude of 80 km is shown in Figure 8 (Silber et al., 2017a). A number of CFD codes with various schemes to treat applications in hypersonic flow exist (e.g., see Hao et al., 2016, and references therein). We further direct the reader to Longo et al. (2007), Gnoffo et al. (2010) and Candler et al. (2015).

A hybrid DSMC-CFD method uses a statistical DSMC method in rarefied flow conditions occurring at high altitudes, coupled with a Navier-Stokes equations solver using CFD in continuum flow conditions at lower altitudes (e.g., Wang et al., 2002; Carlson et al., 2004; Schwartzentruber et al., 2006). A discussion on the formulation of the problem and computational approach are described in Schwartzentruber and Boyd (2006).

Currently there is no comprehensive code to simulate the formation of the shock wave in the rarefied flow during the onset of strong ablation that would be analogous to the CFD and DSMC numerical packages that deal with hypersonic re-entry vehicles gas dynamics both in rarefied and continuum flow regimes.

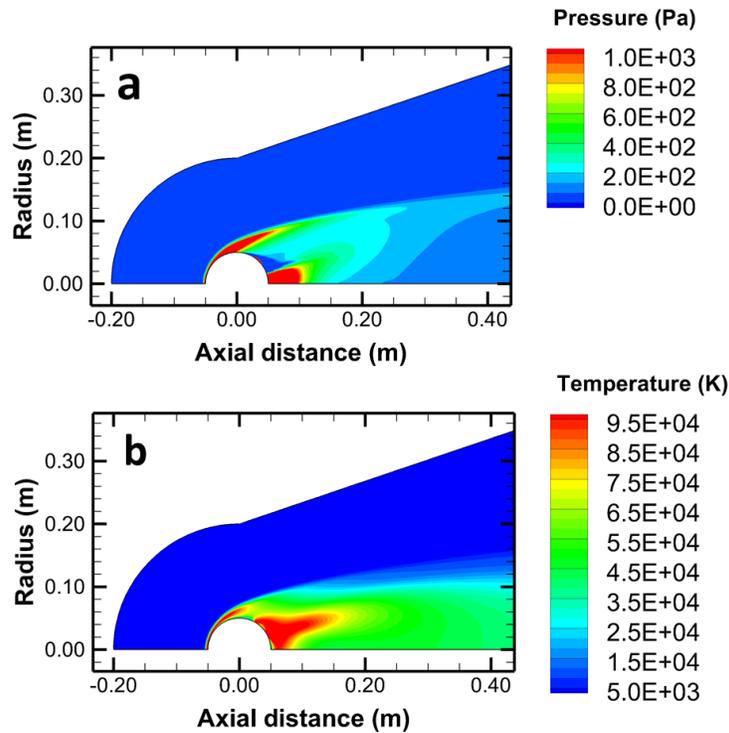

**Figure 8:** The (a) pressure and (b) temperature field around a meteoroid moving at 35 km/s at altitude of 80 km. The diameter of the meteoroid is 10 cm. This example does not include ablation. Credit: Figures 4b and 5b from Silber et al. (2017a).

## 8. Meteoroid Fragmentation

In this section, we reflect briefly on the meteoroid fragmentation phenomena in the context of shock waves. However, due to the complexity associated with the fragmentation and the existence of significant uncertainties, the scope of our discussion is limited, as comprehensive coverage of the topic would require a separate review.

Disintegration of meteoroid bodies during the hypersonic flight takes place when dynamic pressures exceed the binding strength of the meteoroid (e.g., Chul et al., 2012; Register et al., 2017; Tabetah et al., 2018). This phenomenon in generally known as meteoroid fragmentation and it takes place in the regions of material weakness. While there might be several distinct episodes of fragmentation, from a simplistic point of view, it is possible to recognize two basic fragmentations regimes: 1) continuous fragmentation (some interpretations treat this as an integral part of meteoroid ablation); and 2) sudden fragmentation (sometimes referred to as gross fragmentation or discrete fragmentation) (e.g., Ceplecha et al., 2005; Popova et al., 2008).

Meteor fragmentation is a complex process that involves separation of individual fragments from the parent body under the stress of aerodynamic loading, fragment and debris cloud expansion and its spatial and temporal evolution. However, in the context of shock wave discussion, an in-depth review of the topic of fragmentation of larger meteoroids exceeds the scope of this review because of its complexity. This is mainly because the shock analysis in the frame of reference of meteor fragmentation, including the consideration of individual fragments and vapor and dust cloud, needs to account for the interaction of all objects in the system via interacting and coalescing individual shock envelopes in the flow field.

The fragmentation phenomena are more prevalent in larger bodies (Popova, 2011) because the cohesive strength of natural objects, such as for meteoroids, is inversely proportional to the size, and it scales down as the body size increases (e.g., Weibull, 1951; Hartmann, 1969). It should be noted that a fragmentation episode or a gross fragmentation event also corresponds to the sudden increase in the rate of energy deposition into the ambient atmosphere. Consequently, it can be clearly recognized from observations of the light curves behaviour (essentially, fragmentation corresponds to sudden peaks in light emission) (e.g., Svetsov et al., 1995).

In the broader context of shock wave studies, large meteoroid fragmentation, including the special and temporal evolution of disintegrated system of different size bodies, is of great importance for study of space and near space delivery systems that undergo separation stages while at hypersonic and shock generating flows. One of the key reasons is better understanding of the interaction of shock waves by individual bodies and their effect for hydrodynamic stability of the system. Shock wave coupling and flow field interaction becomes increasingly complex when considered from the reference points of individual large fragmented bodies in both the continuum and transitional flow regimes.

In recent decades, the problem of shock wave interaction from fragmented bodies has received considerable attention (Passey et al., 1980; Schultz et al., 1994; Artemieva and Shuvalov, 1996; Artemieva and Shuvalov, 2001; Barri, 2010b). It was demonstrated both experimentally and numerically that when smaller fragments are enveloped by the shock wave of the main largest fragment, smaller fragments may experience a pull toward the axis of travel of the larger body (Schultz and Sugita, 1994; Artemieva and Shuvalov, 2001; Barri, 2010b; Laurence et al., 2007; 2012). This is sometimes referred to as the collimation effect (Barri, 2010b). More recent experimental results and analytical and numerical treatment of the fragmented system demonstrate the strong effects of the size ratio between the larger primary and smaller secondary bodies on the behavior and lateral motion of the latter ones entrained in the main flow field (Laurence et al., 2007; 2012). In principle, the authors show that the larger the size of secondary smaller fragments, the more likely it is that it will be laterally displaced outside of the collective shock envelope bound flow field.

It is important to emphasize that when the fragmentation produce large fragments of similar size, the primary mode of interaction of those fragments is through their shock waves (e.g., Artemieva and Shuvalov, 2001). These fragments subsequently separate further, and might undergo additional fragmentation (Melosh, 1981, 1989; Artemieva and Shuvalov, 1996; Register et al., 2017). However, the process of shock wave interactions and the resulting effects on a flow field of fragmented bodies at hypersonic velocities is a very complex topic and it still remains to be better constrained.

Thus, it can be seen that the problem of simulating the dynamics of meteor fragmentation is not trivial as it needs to account for the combined contribution of individual fragments, dust and small debris, and ablated vapor to the energy deposition in the flow field (e.g., Wheeler et al., 2017). Currently, numerical simulations of meteoroid fragmentation capable of addressing the issues discussed above can be divided into three categories, mainly distinguished by the treatment of the fragment interaction and wake behavior after the breakup. This depends on whether the collective or discrete shock envelope is assumed in the model.

1. *Liquid drop models*. This approach treats a system of fragments, dust particles and vapor as a strengthless liquid (the approach is known as the pancake model) (Zahnle, 1992; Hills et al., 1993; Chyba et al., 1993; Crawford, 1997; Boslough et al., 1997; Shuvalov et al., 1999). The modeled body is permitted to spatially change into the "pancake" only at specific flow regime conditions. This model is useful in describing continuous fragmentation, meteoroid deformation due to the aerodynamic loading and the collective flow field that also includes the initially ablated vapor and plasma that surrounds the fragmented bodies. Subsequently, it is better suited to describe energy deposition in the atmosphere by the expanding and evolving fragmenting system. A common shock envelope assumption is valid when the radius of the dust and ablated vapor cloud is less than two diameters of the initial bolide (e.g. Register et al., 2017).

2. *Discrete fragment models*. These models simulate separate fragments which may or may not interact via shock waves depending on the initial assumptions (e.g., Register et al., 2017). This approach considers the dynamical evolution of a finite number of larger fragments (e.g., Passey and Melosh, 1980) in which case the hydrodynamic approach is not valid (Artemieva and Shuvalov, 2001). Depending on the application, there are three discrete fragmentation models which may be implemented with the collective wake, non-collective wake, and independent wake (Register et al., 2017). This makes it necessary to apply appropriate considerations in terms of fragment interactions via shock waves. In terms of energy deposition, this model is not as efficient as the previous one, because it may not be able to account for the effective drag area generated by vapor and dust particles. The main reason is that the total cross-sectional size of the flow field that includes all fragmented bodies, dust and ablated vapor depends on the degree of assumed mutual interaction of shock waves from individual fragments. Additionally, the size and behavior of the flow field strongly depends on whether the shock envelope encompasses both primary and secondary fragments (e.g., Laurence et al., 2007; 2012)

3. *Hybrid models* (e.g., Register et al., 2017; Wheeler et al., 2017). These models combine the advantages of both discrete fragmentation and hydrodynamic approach to account for the flow field evolution and energy deposition by individual fragments, dust particles and ablated vapor and plasma. Essentially, hybrid models enable a more realistic reproduction of the natural fragmentation phenomenon. As shown by Register et al. (2017) and Wheeler et al. (2017), this type of hybrid fragmentation model is better suited to reproduce the energy deposition and light curves from well documented events, such as Chelyabinsk.

In the context of reproducing natural events, Tabetah and Melosh (2018) implemented a two-material computer code (KFIX) to investigate the fragmentation of objects such as Chelyabinsk at aerodynamics pressures which are about two orders of magnitude lower than the compressive strength of the body. Simulating the exchange of energy and momentum between the meteoroid and the incident atmospheric air, these authors were able to identify a previously unrecognized process in which the penetration of high-pressure vapor and gas from shock layer into the body of the meteoroid enhances the deformation and facilitates the breakup of meteoroids similar to the size of Chelyabinsk. However, modeling challenges still persist in incorporating and reproducing, for instance, better constrained radiative transfer and realistic ablation rates, which remain to be addressed in more sophisticated models.

In conclusion, it is important to emphasize that fragmentation of smaller meteoroids cannot be considered in the same way as fragmentation of large events. In the category of smaller meteoroids (millimeter to centimeter size range), while the initial fragmentation episode may be accompanied by the shock wave, following the separation of fragmented particles, it might not be possible to maintain the collective flow field (within the continuum flow regime) bound by the same shock envelope or even individual particle shock envelopes. This is because the fragmented particles are too small and have proportionally reduced ablation, and as such cannot maintain the continuum flow regime in their flow fields, and consequently cannot generate a shock wave in the MLT.

## 9. Airbursts and NEO Threat

Let us now turn our attention to a larger class of extraterrestrial objects and the shock waves they produce. Asteroids and comets with a perihelion of 1.3 AU or less, thus in the near-Earth environment and capable of crossing its orbit, are collectively known as Near-Earth Objects (NEOs). The sizes of these objects exhibit a power law distribution across many orders of magnitude (Harris, 1998; Brown et al., 2013; Boslough et al., 2015a). Quantification of the distribution comes from many lines of complementary and overlapping evidence. For the smallest objects, the best constraints are from ground-based observational campaigns of meteors (Halliday et al., 1996) and lunar impact monitoring (Suggs et al., 2014). Distributions of somewhat larger objects come from US government sensor data on bolides (Brown et al., 2002) and infrasound data (Silber et al., 2009). Astronomical surveys (Harris, 1998; Mainzer et al., 2011; Tricarico, 2017; Trilling et al., 2017; Stokes et al., 2017; Waszczak et al., 2017) are used for the largest objects. The size distribution is smooth and has no sharp demarcations (Figure 9). Thus, the terms we use for objects and events are somewhat arbitrary and experts in different fields do not always use the same words for the same phenomena. Up until now, this review has primarily been concerned with meteors due to objects smaller than a meter in diameter, which are defined by the International Astronomical Union (IAU) as "meteoroids" (IAU: www.iau.org).

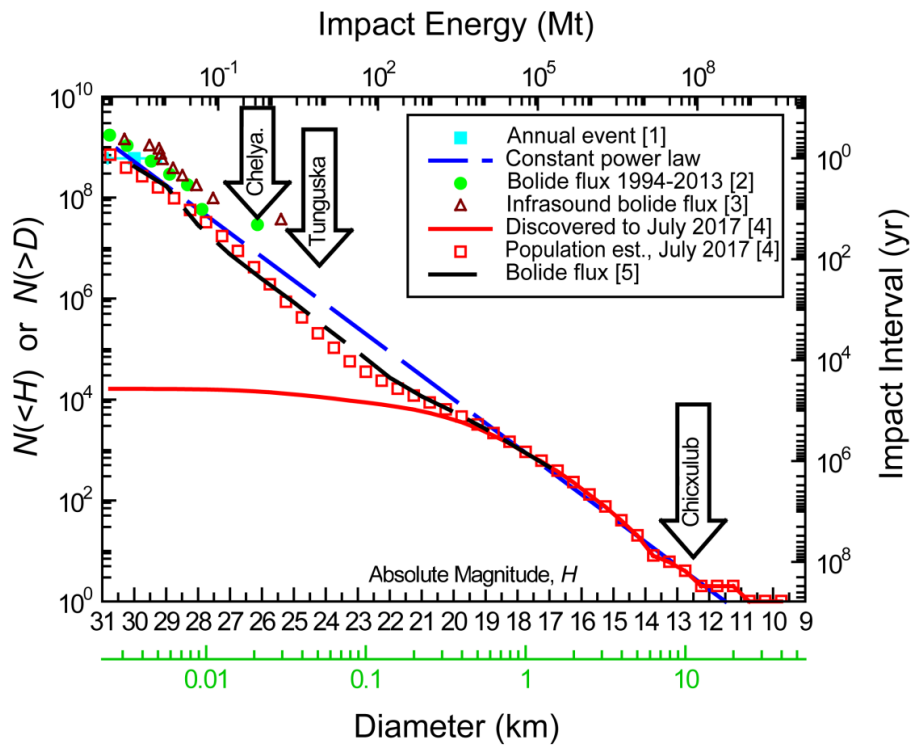

**Figure 9:** Cumulative NEO population estimate from various surveys. The numbers in brackets correspond to the following references: [1] Brown et al. (2002); [2] Boslough et al. (2015a); [3] Silber et al. (2009); [4] Stokes et al. (2017); [5] Tricarico (2017). "Chelya." in the arrow on the far left stands for Chelyabinsk.

Moreover, in this review, we use the word "meteor" as defined by the IAU: "...*the light and associated physical phenomena (heat, shock, ionization), which result from the high speed entry of a solid object from space into a gaseous atmosphere.*" Meteors brighter than magnitude -4 are also called bolides or fireballs, and those brighter than -13 are sometimes referred to as "superbolides". Large superbolides with explosive yields comparable to or greater than weapons of war have come to be known as "airbursts" and are caused by objects greater than about a meter in diameter that enter the atmosphere several times per year. This term has long been used to describe the atmospheric effects of explosive weapons, including nuclear devices. In the words of Glasstone and Dolan (1977): '*Provided the nuclear explosion takes place at an altitude where there is still an appreciable atmosphere, e.g., below about 100,000 feet, the weapon residues almost immediately incorporate material from the surrounding medium and form an intensely hot and luminous mass, roughly spherical in shape, called the "fireball."  An "air burst" is defined as one in which the weapon is exploded in the air at an altitude below 100,000 feet, but at such a height that the fireball (at roughly maximum brilliance in its later stages) does not touch the surface of the earth.*'

Small asteroids of this size disintegrate and mostly or entirely vaporize in the upper atmosphere, causing a high-altitude airburst with an explosive energy on the order of a kiloton of TNT equivalent (1 kt TNT = $4.184 \cdot 10^{12}$ J). These are often meteorite-producing events (e.g., Jenniskens et al., 2012). At the other extreme, large asteroids reach the surface at hypervelocity without losing significant mass to ablation, creating an impact crater (e.g., Brown et al., 2008). Asteroids with diameters of about 20 meters can penetrate into the lower atmosphere (the lower stratosphere or troposphere), where they explode with yields on the order of a megaton (Mt) (1 Mt TNT = $4.184 \cdot 10^{15}$ J). This is roughly the size threshold for NEOs that can cause destruction or casualties at the surface and contribute to the overall impact risk (e.g., Harris, 2018), as demonstrated by the Chelyabinsk event (Brown et al., 2013; Popova et al., 2013). The Chelyabinsk object was ~19 m in diameter, derived from its known density and inferred mass of about 12,000 metric tons. The latter is based on the bolide's known entry velocity of ~19 km/s and estimated explosive yield of ~500 Mt (which is the same as its initial kinetic energy) (Brown et al., 2013; Popova et al., 2013). Denser, faster, or larger objects will have higher yields (with linear, square, and cube scaling, respectively). Because of the power-law size distribution of NEOs, low-altitude airbursts are far more likely hazards than crater-forming impacts and are therefore of intense interest to the planetary defense community (e.g., Mainzer, 2017; Stokes et al., 2017).

This section describes the phenomena associated with the collision of NEOs with sizes between the two bounding cases: 1) complete ablation in the upper atmosphere and no significant effect at the surface, and 2) insignificant fractional mass loss or deceleration before surface impact, with most of the kinetic energy leading to explosive crater formation and associated phenomena. This range of conditions does not depend only on size, but on entry speed and angle, density, strength, and shape of the asteroid. When the conditions are right for an airburst, the asteroid moves up an exponential density gradient as it descends through the atmosphere, leading to dynamic pressures that exceed the material strength and causing it to deform and break. As it breaks up into fragments and changes shape, it continues to descend at hypervelocity into denser air, leading to a rapid increase in drag force and ablation boosted by an increase in its surface area and growing volume of high-temperature plasma. This causes a cascading feedback of mutual reinforcement and abrupt expansion of a multi-phase, turbulent mixture. Because the radiative heating and ablation of fragments surrounded by plasma can take place on a timescale that is short relative to the rate of expansion, the internal pressure can rise abruptly, boosting the explosive expansion of the mixture of air and meteoritic vapor into the low-pressure ambient atmosphere. The energy release is so rapid that it is sometimes referred to as a "detonation". This term is usually reserved for supersonic chemical explosions in which a rapid exothermic reaction reinforces and is reinforced by a shock/detonation wave. It would be a misnomer unless the process includes a self-reinforcing internal shock wave that aids in the release of energy by a means analogous to non-chemical fuel-coolant interaction explosions (e.g., Witte et al., 1970; Corradini et al., 1988). Models of the terminal phases of energy release leading to an airburst have not yet included the detailed physics of such a process.

For purposes of this section, we focus less on the physics of energy release, which is modeled by others (e.g., Shuvalov et al., 2002; Wheeler et al., 2017), and more on the subsequent evolution of the explosion and its effects at the surface which are relevant to the NEO threat. Many published models ignore the actual mechanism of energy conversion because they are unable to address processes on spatial scales significantly smaller than the diameter of the asteroid. Parameterized or semi-analytical models are used to calculate the energy deposition rate along the entry path and to estimate the altitude of maximum energy deposition and how it depends on characteristics of the asteroid. The approach is similar to macroscopic, homogeneous models for chemical explosive detonation. Such models do a good job at predicting phenomena such as plate acceleration or blast generation, but ignore the small-scale physical processes of the detonation itself. To quantify the large-scale effects of a chemical detonation, we can make use of empirically-derived data on thermochemical parameters, material properties, and chemical kinetics of explosives. For this purpose, we can disregard the shear heating, pore-collapse, grain-boundary interactions, and processes that we know control the detonation on spatial and time scales that are small compared to the effects of interest. The effects of airbursts on the ground can be treated with the same philosophy.

Even though they should occur much more frequently than crater-forming impacts, low-altitude airbursts are rare events. Prior to the 2013 explosion over Chelyabinsk, Russia, the only witnessed case was the 1908 Tunguska airburst. For many years, Tunguska provided the only data to compare to airburst models, and on which to base the contributions of airbursts to the impact hazard. The size of the Tunguska airburst is uncertain, and estimates have varied widely. The most widely-quoted magnitude range has been between 10 and 40 Mt (Vasilyev, 1998) and were derived from the nuclear weapons effects literature, based on a combination of historic barometer data, seismograms, and forest damage assessments. Turco et al. (1982) published the highest value, about 700 Mt. The lowest estimate of 3 to 5 Mt was published by Boslough and Crawford (1997). Prior to that, the consensus had placed the yield at the low end of the range of published values – at 10 to 15 Mt – in part because of the unlikeliness of a larger event having taken place so recently. Estimates vary for the average recurrence intervals for Tunguska events. According to Brown et al. (2002), it is about once every millennium for a 10 Mt event. Boslough et al. (2015a) used revised population estimates of Harris et al. (2015) along with revised Tunguska size estimate (~4 Mt) of Boslough and Crawford (1997) to conclude that it was a 500-year event. Harris (2018) now estimates that the flux of objects this size is even lower, perhaps by a factor of two, based on the rapidly-growing catalog of small asteroids. The low-end yield estimates of Boslough and Crawford (1997) had the advantage of being the least unlikely, in terms of size, and were accepted by many researchers.

Observational data that have been used to estimate the Tunguska yield include: 1) areal extent and pattern of fallen trees, 2) seismometer records, 3) barometric pressure measurements, and 4) areal extent of trees charred by thermal radiation. They are constrained by nuclear tests, laboratory experiments, and numerical models. Glasstone and Dolan (1977) derived scaling laws from the effects of nuclear weapons, providing Tunguska researchers with a means to determine the energy release required for a point-source explosion to generate observed phenomena. The pancake model was used by Chyba et al. (1993) to estimate the range of impactor size, strength, and entry speed and was consistent with a 10 to 15 Mt event. This estimate was based on the assumption that the Tunguska explosion had the same effects on the surface as a point-source explosion at the altitude of maximum energy deposition. One justification for the point-source assumption was that observations and models are both consistent with a sharply-peaked energy deposition profile that is concentrated within an atmospheric scale height (Chyba et al., 1993). However, the downward advection of the explosion was neglected. The energy deposition profile is sharply peaked due to mutually reinforcing effects of deformation and drag. This peak was called the "airburst height" (Chyba et al., 1993) and taken as the altitude at which a point source explosion of the same yield would have a similar effect on the ground. However, this assumption neglects the fact that the mass of the projectile is still traveling downward at a significant fraction of its initial speed at the time of its maximum energy loss, and can approach or come into contact with the surface.

Hydrocode simulations of Boslough et al. (2008) revealed that a completely vaporized asteroid inserted at the altitude of maximum energy deposition - but still maintaining the commensurate downward momentum - continues its rapid descent to a lower effective burst altitude. Combining this result with a reexamination of the effects of topography and forest health led to the conclusion that previous analyses had overestimated the yield of the Tunguska event. Taking all these factors into account, Boslough and Crawford (2008) argued that the yield could be 5 Mt or lower (Figure 10).

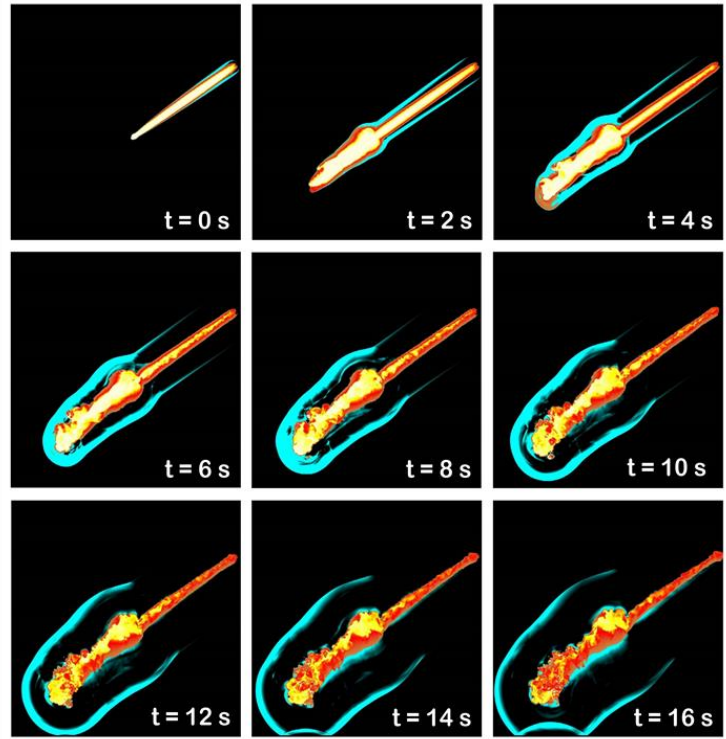

**Figure 10:** CTH simulation time series showing a 5 Mt airburst originating at 12 km above the surface. The impact angle is 45°. The time steps for each frame are shown in the lower right corner (Boslough and Crawford, 2008). The full color figure is shown in the electronic version of the article.

Much of our current understanding of the Tunguska event comes directly from models of the 1994 collision of Comet Shoemaker-Levy 9 (SL9) with Jupiter. Jupiter has no solid surface so the only possible result of an impact is an airburst. Each of the cometary fragments descended to an altitude at which the atmospheric density was high enough that aerodynamic pressure exceeded the fragment's strength. It broke up catastrophically leading to an exponentially reinforcing feedback between the radial expansion and fragmentation, creating increasing drag and dynamic pressure. Hydrocode models showed that when they reached their altitude of maximum energy deposition, the fragments experienced stresses and strain rates of a magnitude similar to those associated with a hypervelocity crater-forming impact onto a solid. The entry wake creates a low-density, high-pressure channel from the depth of the explosion to the top of the atmosphere, leading to a highly anisotropic explosion that is directed upward and outward. Hydrocodes showed the ejection of a high-velocity plume backward along the channel (Boslough et al., 1994b, 1994a) which were observed by the Hubble Space Telescope to rise to 3,000 km above the cloud tops (Hammel et al., 1995). The observational verification of SL9 predictions lent credence to similar hydrocode models for the Tunguska airburst. For example, Tunguska simulations by Boslough and Crawford (1997) yield plumes that rise hundreds of kilometers into space, explaining the high-altitude noctilucent "bright night" clouds and suggesting that low-altitude airbursts could be a hazard to satellites in low-Earth orbit. However, more recent and detailed modeling of Jupiter impacts suggests that there is a level below which upward flow is limited and the plume is "pinched off" (Palotai et al., 2011) and that instabilities along the entry wake may stunt the formation of plumes for impacts as small as the Tunguska event (Pond et al., 2012).

The case for a smaller Tunguska is not solely based on the lower effective height of burst. Boslough and Crawford (1997) also argue that yield estimates based on tree fall are too high because they do not take topography or forest health into consideration. The weapons scaling of Glasstone and Dolan (1977) apply to coniferous forests on flat terrain. Many accounts of the Tunguska aftermath claim that trees were blown down over an area of 2,000 km$^2$, but photographs reveal that many trees were left standing. Slopes greater than 15º are typical within the area of treefall, leading to concentrations of blast wave energy and wind fall extending further than would be experienced over flat terrain (Boslough and Crawford, 2008). There may also be a selection bias toward the most dramatic images of fallen trees, which tend to show ridges where blast energy is concentrated.

Moreover, none of the tree-fall-based yield estimates consider the pre-impact condition of the forest. Florenskiy's (1963) description of his 1961 expedition stated that "*the region of the forest flattened in 1908 was not one of homogeneous primeval intact taiga*," and that "*…the region of meteorite impact in 1908 was basically a fire-devastated area… a partly flattened dead and rotting forest was standing in this area…*" Furthermore, "*…an estimate of the force of the shock wave that is based on the number of flattened trees must necessarily take into consideration the condition of the forest at the time*." Boslough and Crawford (1997) scaled the wind speeds to be consistent with Florenskiy's dynamometer measurements, finding that the necessary point-source yield is reduced to 3.5 Mt.

The bright noctilucent cloud above Europe and western Asia in the nights following the Tunguska explosion may represent independent evidence for a collapsed plume, analogous to similar phenomena observed on Jupiter (Hammel et al., 1995) as described by Boslough and Crawford (1997) and supported by simulations of Artemieva et al. (2007). This view is gradually replacing a long-standing notion that the bright nights are attributable to the tail of an impacting comet (e.g., Bronshten, 2000).

Boslough and Crawford (2008) ran a new set of Tunguska-scale airburst simulations in anticipation that significant impact energy is transported downward within the exploding fireball. The main motivation was to develop a qualitative understanding by exploring parameter space, leading to new models to more properly capture the physics. They varied size, impact energy, and material properties (strength and density) over a wide range of values in dozens of simulations using adaptive mesh refinement to resolve both the asteroid and the entire domain that experiences the effects of the airburst. These simulations revealed unexpected but robust phenomena that emerged as a result of a large range of realistic assumptions.

Because the effects at the surface are dominated by post-airburst hydrodynamics, the simulations were simplified by adding an additional internal energy source term to initiate the explosion at a prescribed altitude. They sourced 2.0 Mt into an asteroid with 3.0 Mt of kinetic energy, in order to provide a bounding case for a 5.0 Mt airburst. The high energy density after the energy insertion causes the asteroid to vaporize immediately after completion of the artificial energy insertion, leading to instant loss of strength and explosive expansion of a fireball-like jet which maintains its downward momentum. This extreme case was intended to provide a computationally-based bound on the phenomena generated by rapid asteroid fragmentation, and ablation. One justification for a quasi-instantaneous conversion of kinetic to internal energy of a Tunguska-scale asteroid is based on simulations by Svetsov (1996) suggesting that a stony asteroid can be broken by aerodynamic force into 10 cm and smaller fragments that are quickly ablated by the high-temperature fireball in which they are immersed. The hot jet of air and asteroid vapor continues its hypersonic descent driving a downward bow shock ahead of it, which is reinforced by the explosive expansion. This generates a single wave that is simultaneously fed by both translational and radial components of kinetic energy, in contrast with descriptions in some of the Tunguska literature, in which a "ballistic wave" (the bow shock) and "explosion wave" (radial shock) are described as two distinct phenomena (e.g., Zotkin et al., 1966).

Other robust phenomena that emerged are large toroidal vortices that are initiated by the rapidly descending hot jet. Ring vortices greatly reduce the aerodynamic drag on the expanding fireball, which means it can descend to a much lower altitude before stopping than if it were a rigid object. The simulations suggest that vortex formation may be a consequence of chaotic perturbations in the large-scale hydrodynamic flow, implying that large-scale effects on the ground may depend on a non-deterministic fragmentation process that controls macroscopic flow. This chaotic effect would lead to statistical uncertainties in the hazard associated with low altitude airbursts. Asteroid entry and fragmentation models must therefore include the "post-burst" phase that takes into account the reduced drag on a rapidly expanding and descending fireball.

In some simulations, the fireball descends to altitudes much lower than that at which the asteroid explodes, but does not reach the surface. When its rapid descent stagnates, it reaches a minimum altitude and then rises buoyantly, like a nuclear mushroom cloud. All else being equal, a larger NEO will explode at a lower altitude and its fireball will come closer to the surface, leading to stronger blast waves and more intense thermal radiation. Above some size threshold, the descending fireball will make contact with the surface and damage within the contact zone will be much more extreme due to the much higher temperatures and wind speeds within the fireball. For example, depending on the energy of the bolide, wind speeds resulting from the blast wave can attain velocities high enough to cause significant damage in the affected area (Table 1) (Glasstone and Dolan, 1977). Boslough and Crawford (2008) suggested that such a surface-contacting airburst can melt surface materials which can then quench to form glass. Such a mechanism would explain the 29-million-year-old Libyan Desert Glass, an enigmatic pure silica glass found in the western desert of Egypt. Boslough and Crawford (2008) suggested classifying airbursts as "Type 1" for the Tunguska event and "Type 2" for the putative Libyan Desert Glass event (below and above the size threshold for surface contact, respectively).

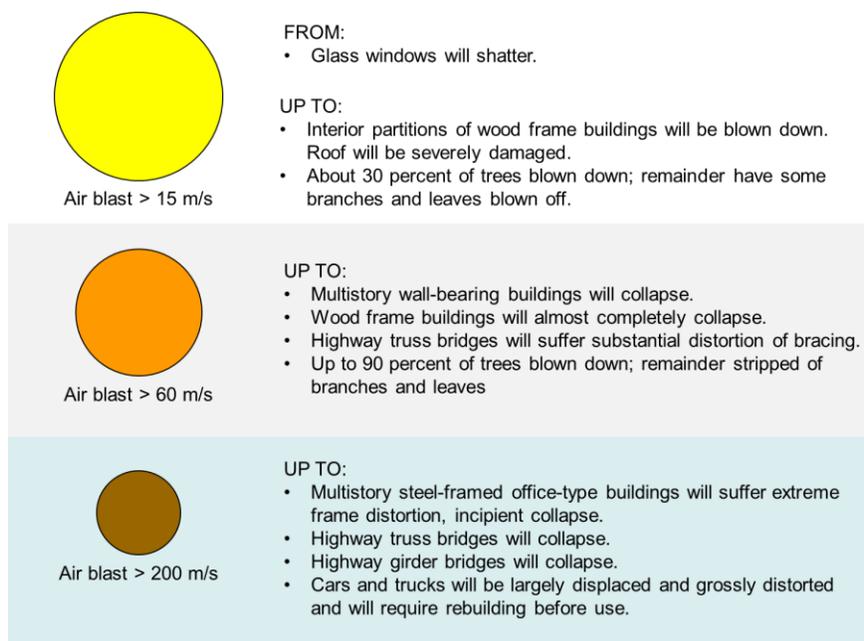

**Table 1:** A summary and breakdown of the extent of damage resulting from an air blast (for further details, see Glasstone and Dolan, 1977; Collins et al., 2005; the simulator is available at: https://www.purdue.edu/impactearth).

Asteroid-generated tsunami are also potential contributors to the overall risk. Boslough (2013) suggested that airbursts might enhance the coupling efficiency due to plume ejection and air blast though a mechanism similar to that which creates meteotsunamis on Earth and the large atmospheric waves observed on Jupiter after the impact of comet fragments (Hammel et al., 1995). Preliminary modeling to test this idea is interpreted by Stokes et al. (2017) to suggest that insufficient energy is coupled from the air blast to a shallow water wave to represent a significant contribution to the NEO risk. They argue that because most of the shock wave energy is reflected from the ocean surface in the short time-frame of the simulations, coupling could be neglected. However, much longer simulations over much larger domains will be required to address other suggested coupling mechanisms, including Proudman resonances with radial rarefaction waves, and momentum enhancement from plume ejection and collapse.

Computational models for airbursts are useful both for gaining insight into the physics and for generating quantitative damage estimates. The emergent phenomena (e.g., ballistic plumes, downward jets, and large-scale vortices) that were discovered by the hydrocodes must be factored into impact risk assessments and can guide the search for evidence in the observational, historical, and geologic record. Parameterized and semi-analytical models can be modified to include vortex formation, buoyant forces, and surface interactions of the fireball. Air blast and radiative damage estimates can be improved across the span of possible airburst scenarios. Calculated damage maps based on hydrocode simulations can be compared to the actual data to better quantify the low-altitude airburst at Tunguska. Better risk assessments can be based on these simulations by combining damage estimates with impact probability density functions to inform policy decisions for planetary defense.

Low-altitude airbursts are by far the most frequent impact events that can cause damage or human casualties on the ground. It can always be said that the next impact with such consequences on Earth will almost certainly be another low-altitude airburst. Figure 11 shows the density map as a function of wind speed for bolides delivering energy between 0.5 and 16 Mt. In comparison to a typical point source explosion for a given yield, airbursts are far more damaging, as illustrated in Figure 12. The primary purpose of airburst simulations thus far is to gain insight about the phenomenon and to develop assessment and mitigation methods of the potential threat (for a more comprehensive review, see Artemieva et al., 2016). We now have the computational tools capable of performing large ensembles of simulations of asteroid airbursts over a range of size, entry speed and angle, and burst altitude. We can generate damage maps on the ground to estimate the costs associated with infrastructure and economic losses. This capability has been used to support tabletop exercises for the US Federal Emergency Management Agency (FEMA) (Boslough et al., 2015b) and threat exercises as part of the biannual Planetary Defense Conferences (Boslough et al., 2016). Computational fidelity of such simulations can be enhanced further by adding more physics and capability to existing hydrocodes, and by using more powerful computers and methods to extend simulations to smaller scales that resolve the cascade of fragmentation and ablation that leads to the explosion and subsequent phenomena.

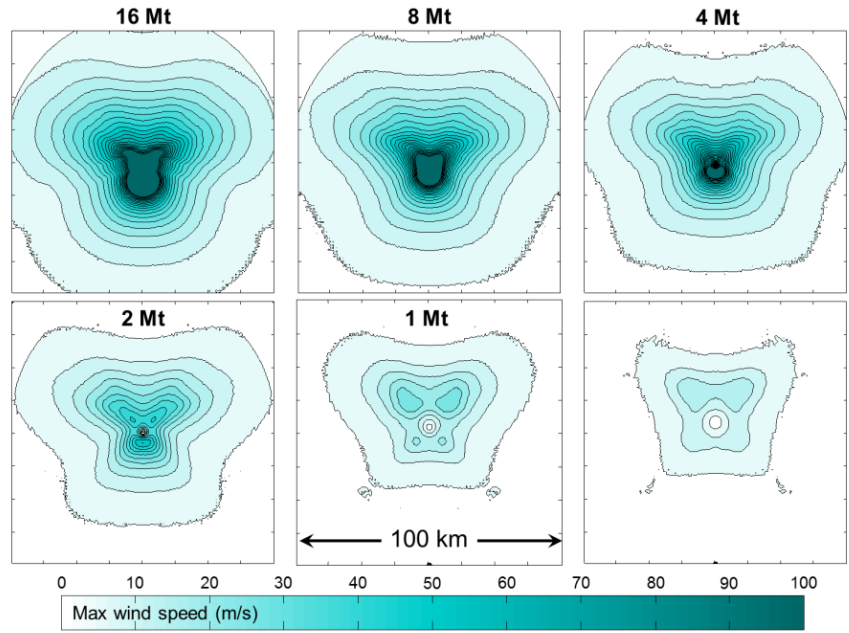

**Figure 11:** Density maps (ground "footprint") showing wind speeds resulting from airbursts originating at ~12 km altitude. Entry angle for all cases is 45°, with the direction of arrival from north (top) to south (bottom of the figure). The energy delivery ranges from the highest value of 16 Mt down to 0.5 Mt, as shown in each panel. The wind speed is in m/s. Each panel is 100 km wide.

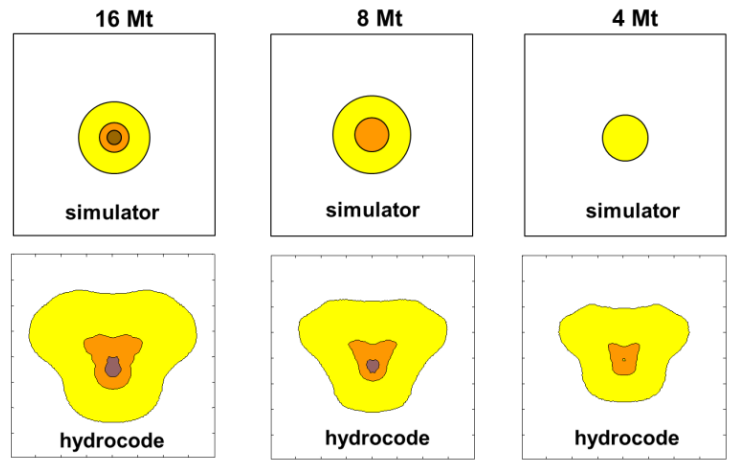

**Figure 12:** The wind speed density (ground "footprint") distribution and area affected by a deep penetrating bolide delivering 16, 8 and 4 Mt of energy (from left to right). The shading corresponds to the three categories shown in Table 1. The air blast speed is as follows: darkest shading (brown) is > 200 m/s, medium shading (orange) is > 60 km/s and lightest shading (yellow) is >15 km/s. The upper row shows the estimates using point source relations (Collins et al., 2005; the simulator is available at: https://www.purdue.edu/impactearth). The lower row shows the results from the airburst hydrocode simulations for a bolide traveling from north (top) to south (bottom of the figure).

## 10. Detections of Meteor Generated Shock Waves Using Radar and Infrasound: A Brief Overview

*10.1 Radar*

The contemporary set up and capabilities of modern meteor radar systems do not permit direct detection of the meteor generated shockwaves. However, it might be possible to use meteor radars to observe secondary effects that are associated with, or are indicative of meteor shock wave formation, as discussed by Rajchl (1969, 1972) and Silber et al. (2017a). In that sense, it is important to reflect briefly on the fundamentals of meteor radars.

Studies of meteors with radar can be classified broadly into two extremes which employ quite different strategies. While most meteor radars use pulsed transmitters for range-resolution, they differ widely in many other features. With regard to these features, the first class are large high-power radars (referred herein as HPLA) which are commonly used to study meteoroids as they enter the radar beam (e.g., Chau et al., 2004; Westman et al., 2004; Close et al., 2007, 2008; Janches et al., 2008). These radars use high power, large collecting areas and narrow beams, and are designed to receive reflections and/or scatter from the head of the meteor (e.g., Pellinen-Wannberg, 2005; Dyrud et al., 2008) - specifically from the high-density plasma vapor cloud discussed earlier (Hocking et al., 2016a). Numerical and theoretical studies have been performed to simulate the signal received (e.g., Räbinä et al., 2016; Marshall et al., 2017; Dyrud and Janches, 2008). However, few such studies have looked at the impact of shock fronts on the radio-waves, and there is considerable extra work to be done in this area. Recently, the possible relation between the termination of meteor head echoes, detected by HPLA (High-Power, Large-Aperture) radars, and the onset of meteor generated shock waves has been discussed by Silber et al. (2017b). However, further theoretical work and experimental validation is required to confirm the initial conclusions of that theoretical reasoning.

The second extreme class of meteor radar does not use high power or narrow beams. Nor do such radars receive signal reflected primarily from the meteoroid head, but rather from the entire ionized meteor trail left behind by the meteoroid. These radars rely on quite broad beams - often almost isotropic. They generally operate at VHF frequencies, typically in the range 15 to 50 MHz, whereas the HPLA class often use frequencies of hundreds of MHz and more. This means the HPLA may have wider bandwidths and hence better resolution, often approaching a few tens of meter, in contrast to range resolutions of typically 2-3 km with the radars being now discussed (Hocking et al., 2016a). We will refer to this second class of radars as IMR, for Interferometric Meteor Radar (with reasoning to be established shortly) (also see Hocking et al. 2001; Holdsworth et al., 2004; Mathews, 2004; Malhotra et al., 2007; Janches et al., 2014).

The meteor trails require special alignment in order to be detected by IMR, since they rely on so-called "specular reflection" in order to be detected. This means that the spatial orientation of the trail axis (the vector along the direction of meteoroid propagation path) needs to be perpendicular to the vector from the radar to the trail. However, because scatter occurs along the entire length of the trail, and because of the choice of frequencies, the trail offers a large cross-section for the radar, and these types of radars detect far more meteors than most narrow-beam systems (Hocking et al., 2016a). IMR Systems exist which are capable of detecting 30,000 meteors per day at modest power, such as 10-30 kW peak power (e.g., Fritts et al., 2012). This compares to typically Megawatts of peak power for the largest HPLA discussed above.

However, these smaller radars cannot be used in monostatic mode. Because of the large cross-sections of the meteor trail, they can detect to magnitudes far weaker than the larger radars and this causes issues in location. A huge fraction of meteors are detected through sidelobes of the radar beams, as was the case with the Poker Flat Meteor radar (Avery et al., 1983), especially if the main beam points nearly vertically (since meteors are optimally detected at angles of 30 to 60 degrees from azimuth, which coincides with the location of many sidelobes in a system like Poker Flat). In order to compensate for this, it is generally mandatory to receive on multiple receiver antennas and use phase-interferometry between receiver antennas to locate the direction of the received signal (hence leading to the suggestion of IMR for classifying these radars). Radar pulsing is used to determine range, as is normal for all pulsed systems. Detection rates can be further enhanced through pulse-coding techniques. IMR radars are commonly used to determine winds and temperatures in the mesopause region, apart from their ability to teach us about meteors (e.g., Tsutsumi et al., 1994; Singer et al., 2003; Pancheva et al., 2004; Franke et al., 2005; Buriti et al., 2008; also see Hocking et al., 2016a, section 11.3).

A suitably designed IMR system can achieve angular accuracies of 1-2 degrees and altitude accuracies of 2-3 km. Examples of such radars are shown, and the most common details of operation are discussed in Hocking et al. (2001) and also Hocking et al. (2016a), section 11.3. The receivers are not widely spaced - typically they can be enclosed in a square of width 5 radar wavelengths. However, even when using interferometry, some caution is required. It is easy to detect weak meteors which have sufficient phase-noise that it prohibits accurate echo location. If a height accuracy of worse than 4 km is produced, the data-point should really be discarded for many applications. Often more than half of the meteors that can be detected should be eliminated for such resolution reasons. Each meteor measurement should have some form of error determination as part of the interferometric processing. However, it is not uncommon for operators of such radars to solely use count-rates as an indicator of system capability. This can be very misleading - a system detecting say 30,000 meteors per day, but accepting all meteors can appear much more effective than one which detects only 15,000 meteors per day but uses high quality error-checking, yet the latter may well be the more effective system. Suitability often depends on the type of data required.

While we have distinguished HPLA and IMR, there are some systems which employ some aspects of each radar-type. The SAAMER and DRAAMER systems (Fritts et al., 2012) use an interesting combination with 8 distinct beams, combined with interferometry, which allows them to detect at least a few head-echoes daily, as well as perform traditional IMR studies. At times special echoes called "non-specular" echoes are detected by IMR radars, where the scattering seems to be due to plasma instabilities and not specular reflections. These are often seen close to regions of magnetic field lines aligned perpendicular to the radar/scattering-region vector. The scatter is not due to shock fronts and we will not pursue these further here.

IMR usually concentrate on so-called "underdense" meteors. These are defined to be meteor trails with line densities under $2.4 \cdot 10^{14}$ electrons per meter of trail length (e.g., McKinley, 1961; Poulter et al., 1977, 1978; Silber et al., 2017a). Overdense meteors have line densities in excess of $10^{16}$ m$^{-1}$ (Poulter and Baggaley, 1977, 1978; Jones et al., 1990). Meteor trails with intermediate lines densities are referred to as "transitional". The choice of a line density of $2.4 \cdot 10^{14}$ m$^{-1}$ ensures that most radio VHF waves can penetrate through the trail if the line density is below this value. Appendix C of Hocking et al. (2016b) discusses the parameters needed for trail penetration. If the trail is underdense, radio-waves may penetrate the trail and are scattered from electrons throughout the trail. As the trail diffuses outward and widens to a value of the order of half a radar wavelength, radio-wave interference across the trail causes the signal to decrease exponentially with time, so the backscattered signal decays exponentially, dying out in a few tenths of a second. The majority of meteors have this character, and are used for atmospheric wind and temperature determinations (e.g., see Liller et al., 1954; Cervera et al., 2000; Cevolani et al., 2003; Meek et al., 2013). However, such meteor trails are generally created by quite small meteoroids, and so are unlikely to generate shocks.

With regard to this paper, it is the meteors that generate shocks that are of most interest. Transitionally dense and overdense meteors are more likely to generate shocks than uderdense meteors, although there is still uncertainty regarding the specific regime in which shocks are expected to occur, as discussed earlier. In principle, shockwave generating meteors detected by meteor radar are primarily of strong transitional or overdense nature (Hocking et al., 2016b). These more dense trails start their lives as trails with peak electron densities higher than the plasma frequency for most IMRs. The radio reflection from the trail is very strong, like that of a radar signal reflecting from a solid metal cylinder. As time goes by, they spread and weaken, eventually achieving electron densities with plasma frequencies below the radar frequency. At that time they become partially reflecting, but if their width is already wider than typically half a wavelength, the backscattered signals from different parts of the trail destructively interfere and the signal immediately dies out. Lifetimes of these overdense trails can be several seconds, but most IMRs have software-defined limits on the lifetimes of typically 2-4 seconds, to avoid allowing non-meteor signals to be accepted.

A detailed study of the portion of these stronger meteors which could be detected by a typical IMR was presented by Hocking et al. (2016b), appendix C. However, based on likely lifetimes, and to a lesser extent the likelihood that the received signal might saturate typical receivers, it was concluded that even the strongest meteors detected by IMR are probably upper-level transitional meteors, and generally not truly overdense meteors.

Possible applications of IMRs to studies of shock fronts are still under investigation. Most potential applications will be statistical in nature: in contrast to HPLAs, it is unlikely that studies of any individual meteor event will reveal much. Nevertheless, some useful studies have been made. One such example is Hocking et al. (2016b). In this paper, the complementary cumulative number of meteors was plotted as a function of decay time, and the behavior at the large-decay time end (decay times of a second or so) were examined and compared to ozone densities at multiple sites, both for night-time and daytime conditions. The paper presented convincing evidence that ozone plays a key role in hastening the rate of decay of meteor trails. The explanation requires the existence of shock waves.

Some key points from the paper include not only the role of ozone in the plasma trail decay, but the need for care and consistency in analysis. Every meteor radar a uses slightly different way of identifying meteors. Some require an exponential decay, others require only that the signal decays with time. The former method would miss many of the non-underdense meteors, since their decay is not usually exponential - rather, it holds steady for a while and then drops off relatively rapidly as the trail expands and becomes transparent (see above). So it is important that the same radars and software are used for comparisons between sites. In this case all the radars were SKiYMET radars (Hocking et al., 2001).

That paper also demonstrated another important point. It is not uncommon in searching for mechanisms that authors need to decide between different processes. For example, is the process due to diffusion or chemistry? The important, but simple point that emerges from this paper is that the mechanisms do not need be singular. In this case, three mechanisms were involved - shocks, diffusion and chemistry all contributing some aspects. Furthermore, ozone had to diffuse in from the surrounding atmosphere, once the shock-front had passed. All ozone-related chemistry took place on the edge of the meteor trail. As discussed in Silber et al. (2017a), the presence of short duration hyperthermal chemistry on the boundaries of postadiabatically expanding meteor train is indicative of the prior occurrence of the relatively strong cylindrical shock waves. This thermally driven chemistry between the shock modified ambient atmosphere and meteor metallic ions on the boundary of high temperature meteor train, that very rapidly removes electrons, could potentially be detected by meteor radar (Silber et al., 2017a).

Further work like this with both HPLA and IMR will be important in further understanding the complex physical processes involved in the meteor trail formation, especially with regard to the early stages of trail formation and the complex physics and chemistry that take place around the meteor head and early shock-fronts.

*10.2 Infrasound*

In preceding sections we discussed the shock waves produced by hypervelocity meteoroids during their passage through the atmosphere. Here, we will give a brief overview of a byproduct resulting from strong shock decay over a distance – a very low frequency acoustic (or infrasound) wave. Since infrasound is a manifestation of shock waves, it can be used to estimate energy released by meteoroids (e.g., Silber et al., 2009, 2011; Brown et al., 2013), and in some cases other parameters, such as the altitude of the shock (Silber et al., 2014), which is relevant for establishing flow regimes (Moreno-Ibáñez et al., 2018) and testing and validation of shock propagation models (e.g., Silber et al., 2015; Nemec et al., 2017). A detailed overview of infrasound from meteors can be found in ReVelle (1974), Edwards (2010), ReVelle (2010), Silber (2014), and Silber and Brown (2019).

Infrasound is acoustic energy generally considered to be in the frequency range of 0.01 Hz to ~20 Hz, below the low-end of human hearing. The combined effect of atmospheric spreading and absorption losses (e.g., Sutherland et al., 2004) is proportional to the square of the airwave frequency ($f^2$), so that losses are least in the lower frequencies; thus, infrasound can propagate over very long distances with little attenuation. Meteors, volcanoes and earthquakes are examples of natural sources, while explosions and mining activity are common (but not the only) sources of anthropogenic infrasound sources. Infrasound data, recorded typically as pressure-time series or waveforms, are generally recorded by arrays of four or more low frequency microphones placed in a regular planar arrangement on the ground surface. The analog data are sampled and digitized at 20 samples per second, or higher, and stored. Standard array processing techniques are used to process the data to determine the direction of arrival (back azimuth) of coherent plane waves moving across the array. Processing also gives the apparent horizontal velocity across the array, known as the trace velocity as well as some measure of coherence or correlation of the signal energy (e.g., Christie et al., 2010).

The speed of sound ($c$) in the atmosphere is a function of temperature ($T$) and is expressed as: $c = (\gamma RT)^{1/2}$, where $R$ is the specific gas constant. Thus, thermal structure of the atmosphere plays an instrumental role in long range acoustic propagation. The dynamic variability of the atmosphere occurs on timescales of airwave propagation, which might irreversibly affect the signal by the time it reaches the receiver (e.g. by refraction, ducting and reflection) (de Groot-Hedlin et al., 2010). This presents challenges in source geolocation and characterization (e.g., Silber and Brown, 2014). For example, the atmospheric winds can reach a reasonable fraction of the sound speed and thus affect the observed pressure amplitudes, as well as bias the measured back azimuths (e.g., Donn et al., 1971; de Groot-Hedlin et al., 2010; Mutschlecner et al., 2010). Moreover, the seasonal winds in the upper stratosphere (at ~ 50 km altitude) can create a favorable propagation duct when the energy is moving in the directions of those winds (e.g., Tolstoy, 1973; Georges et al., 1977; Drob et al., 2003; Kulichkov, 2010). Because meteoroids can produce shock waves at any altitude up to ~ 100 km and in rare cases even higher (depending on the parameters described in preceding sections, such as $Kn$) (Zinn et al., 2004; Brown et al., 2007; Silber and Brown, 2014), the resulting infrasound signals may take multiple paths through the atmosphere, and even reach altitudes of 120 km before being refracted back to the surface. Thus, the full atmospheric state from the ground to those altitudes is needed for source characterization, although this in itself is a challenging task (e.g., see de Groot-Hedlin et al., 2010).

Meteoroid infrasound production occurs in three ways (Ceplecha et al., 1998): 1) when a strong shock wave formed during the meteoroid's hypervelocity passage through the atmosphere decays into an acoustic signature (sonic boom); 2) when the meteoroid fragments with an explosive release of energy during the atmospheric flight (e.g., ReVelle, 1976; Silber et al., 2011, 2015 and references therein; Brown et al., 2013); and 3) as a result of a surface impact (e.g., Brown et al., 2008). In case 1), the meteoroid moves so fast that the Mach cone can be approximated as a cylinder and thus the object acts a as cylindrical line source (Tsikulin, 1970). In case 2), gross fragmentation can be considered as a point source explosion, or quasi-spherical source (e.g., ReVelle, 1976; 2005). The situation can be further complicated with multiple fragmentation episodes, or where both cylindrical and spherical shocks might occur simultaneously. To conform to the general theme of this review paper, the discussion in this section is mainly focused on cylindrical shock waves. A primer on shock wave theory relevant to meteor infrasound is given by Silber (2014).

The first recorded infrasound signal by a terrestrial body is from the Tunguska event (Whipple, 1930). However, it was not until the Cold War era that infrasound gained popularity as a passive and relatively inexpensive tool in detecting large explosions in the atmosphere. The United States Air Force Technical Applications Centre (AFTAC) operated a global network of infrasound arrays optimized to detect large explosions (e.g., Revelle, 1997; Silber et al., 2009). Although not immediately recognized, some of the explosive sources recorded between 1960 and 1974 by infrasound instruments were produced by large bolides (Shoemaker et al., 1967). Recently digitized and re-analyzed (Silber et al., 2009), this historical data set provides valuable information about the flux of objects meters to 10s of meters in size (also see Figure 9), especially since it covers the time period when observations of extraterrestrial objects were scarce.

Currently, the infrasound component of the International Monitoring System (IMS) of the Comprehensive Nuclear Test Ban Treaty Organization (CTBTO), provides global monitoring for illicit explosions (Christie and Campus, 2010) and contributes to the growing body of infrasound data from other sources, including meteors. Infrasound from larger events (many kilotons of energy release) may be detected at multiple arrays over thousands of kilometers. Notable examples are the 2009 daylight bolide over Indonesia (Silber et al., 2011) and more recently, a spectacular fireball over Chelyabinsk (e.g., Brown et al., 2013).

Energy released by a meteor is of great interest to scientific community, as it is one of the most crucial parameters in constraining the NEO flux and NEO hazard. Source yield can be estimated by analyzing the measured signal attributes, such as the airwave period and/or amplitude observed at an array. There exist a number of empirical relations relating signal amplitude to energy release; however, the most robust estimate is given by empirical period-yield relations, originally developed for nuclear explosions (ReVelle, 1997). A compilation of various empirical energy relations is given in Silber and Brown (2018).

A particular infrasound signal that was commonly observed from large explosions is the Lamb edge-wave or fundamental atmospheric mode (e.g., ReVelle, 2008, 2010 and references therein). This is a horizontally propagating wave along the surface that suffers little attenuation and decays exponentially with height. It would be observed as a long period wave arriving with the surface sound speed, and should be quite prominent. The Lamb waves approach presented by Pierce and Posey (1971) (also see ReVelle et al., 1996) can also be used to estimate explosive source yield. While the Lamb waves were commonly observed in large nuclear explosions, small bolides would not be likely candidates for such waves. ReVelle (2008) developed a Lamb wave prediction model which he then used on four large bolides, including the Tunguska and the Revelstoke events. The latter is one of the large bolides from the historical dataset (e.g., Silber et al., 2009). He also discussed the uncertainties associated with determination of the Lamb wave source, as its manifestation alone might not be indicative of the bolide entry (for further details, see ReVelle, 2008). For further discussion on the Lamb waves and conditions under which they are likely to form, see ReVelle (2010).

Both analytical and numerical approaches have been developed in an attempt to predict an infrasonic signature pattern at the receiver. Drawing upon early works on cylindrical line source related to lightning (e.g., Jones et al., 1968; Few, 1969), exploding wires (i.e., Sakurai, 1964; Plooster, 1970) and meteors, ReVelle (1974) formulated an analytical blast wave (or weak shock) model for meteors. According to this model, the initially formed strong shock generated by the meteoroid decays to weak shock of the N-wave (a typical sonic boom signature that resembles the letter N) at some radial distance, assumed to correspond to $10R_0$, where $R_0$ is the blast or relaxation radius. As mentioned in Section 6, there are several expressions for $R_0$. Although these are fundamentally the same, they differ by some proportionality constant. The list of various expressions can be found in Silber (2014). At $10R_0$, the fundamental signal period ($\tau_0$) can be related to $R_0$ via: $\tau_0 = 2.81(R_0/c_0)$, where $c_0$ is the local ambient speed of sound. This means that meteoroids depositing large amounts of energy will also have large blast radii and consequently a long fundamental period of the wave. Since the frequency is inversely proportional to the period, the larger the event, the lower the frequency, and thus the longer the propagation distance of the resulting infrasonic wave. Before reaching the receiver, the weak shock can 1) continue to propagate in the form of weak shock; or 2) transition into a linear wave at some "distortion" distance. In either case, for the ReVelle's (1974) weak shock model for meteors to be valid, certain approximations have to be made, such as that no fragmentation occurs and ablation is negligible (i.e., non-fragmenting, single body meteoroid). Another condition is that all arrivals must be direct, meaning that infrasound signal cannot reflect or refract between the source and the receiver (distance < 300 km). Using simultaneous optical and infrasound observations of centimeter sized meteoroids detected over a period of several years, Silber and Brown (2014) developed an approach to determine the shock source height and type of shock (cylindrical or quasi-spherical). This high fidelity data set was later used by Silber et al. (2015) to test and validate ReVelle's (1974) weak shock model. The derivation and full treatment of ReVelle's (1974) weak shock approach are outlined in Silber (2014), and Silber and Brown (2019). While ReVelle's (1974) analytical treatment offers a good first-order approximation for direct arrivals, it suffers from inherent limitations (e.g., no ablation and no fragmentation).

The next natural step forward is a numerical approach, although a full treatment for meteor generated shock waves (with ablation) at all altitudes and across all meteoroid velocities is still lacking. A very limited number of studies have been done in the domain of meteor generated infrasound. Aircraft sonic boom theory (e.g., Whitham, 1974; Maglieri et al., 2014) has been recently applied to meteor shocks. The crater forming impact over Carancas, Peru (e.g., Brown et al., 2008) sparked great interest in modeling meteor generated infrasound. Although the detailed meteoroid entry parameters necessary for high fidelity validation and model testing are not known in great detail, this event has been used as a test case in numerous modeling studies (e.g., Haynes et al., 2013; Henneton et al., 2015; Gainville et al., 2017). More recently, Nemec et al. (2017) applied aircraft sonic boom theory to four well constrained meteor events from Silber et al. (2015). Johnston et al. (2018) developed a coupled radiation and ablation model to describe the aerothermodynamic environment of atmospheric entry for large meteoroids (1 – 100 m) moving at velocities from 14 to 20 km/s at altitudes of 20 – 50 km.

Although all these studies yield promising results, much more work is needed in this domain. Future studies should aim to extend the model capabilities to higher velocities and altitude, with the inclusion of ablation treatment.

**11. Summary and Conclusions**

Shock waves are generated by almost all strongly ablating and sufficiently large meteoroids that enter denser regions of the Earth's atmosphere and reach the lower transitional flow regime as defined by the Knudsen number (provided that these meteoroids retain sufficient mass and velocity close to that of the initial pre-atmospheric velocity). Moreover, based on the mass law, the overall number of meteoroids capable of shock formation is only a fraction of the overall meteor influx. However, the problem of explaining meteor generated shock waves is more complex and differs from analogous shock waves produced by much larger and weakly ablating hypersonic re-entry vehicles. The main reason for such difference is the much higher propagation velocity of an average meteoroid and the presence of ablation induced hydrodynamic shielding that alters the flow regime and pushes the continuum flow further up in altitude relative to the actual meteoroid dimensions.

While the shock in front of the meteoroid dissociates, excites and ionizes the volume of the initially swept atmosphere, the presence of high kinetic temperatures behind the shock front, coupled with strong ablation, provides the background conditions for generation of meteor generated cylindrical shock waves as if they were formed by an explosive line source. The strength of those cylindrical shock waves is also a direct function of the energy deposition by the ablating meteoroid per unit path length.

Analytical and numerical approach in the rarefied flow is difficult, and more so in the case of strong ablation. Some fundamentals of the Eulerian and Navier-Stokes equations have been discussed, along with an overview of the current numerical methods dealing with the rarefied flow. However, currently there is no comprehensive numerical model that can account for the shock generating meteor flow fields when strong ablation in the regime corresponding to a high Knudsen number is present.

Deep penetrating meteoroids can produce airbursts, which can have destructive consequences at the ground, even if a physical object does not make a touchdown. Considering that only a couple of such events have been documented, much of our knowledge comes from numerical studies. Numerical modeling is also the key in evaluating planetary threats from NEOs.

The direct detection of meteor generated shockwaves in the upper atmosphere is difficult because of the bright luminous phenomenon that accompanies the shock passage and the rapid shock attenuation. In that respect, we have reviewed some basics of meteor radar science and infrasound. The radar methodology is capable of detecting the indirect effects of meteor generated shock waves that are materialized through the presence of short lasting hyperthermal chemical reactions and subsequent rapid initial electron removal from boundaries of diffusing meteor train. The infrasonic methodology can be used to determine energy deposition by a meteoroid, and in some circumstances, the shock source height above the Earth's surface. Both methodologies may be promising for improved meteor shock wave detection and observation.

We conclude by proposing that meteor science is not a science in decline. That is evident as there are still unanswered questions, especially in the area of meteor shock waves. Thus, much more work remains to be done. Some of the questions that remain to be answered are:

1. Resolving the altitudes where cm-sized meteoroids generate shock waves as a function of velocity, size and composition (Moreno-Ibáñez et al., 2018).
2. Understanding the physico-chemical aspects of meteor shock wave phenomena in the near and far field ambient atmosphere around the meteor and its physico-chemical impact on the mesosphere and lower thermosphere (MLT) (e.g., Vadas et al., 2014; Silber et al., 2017a).
3. Possible methods for meteor shock wave detection at the altitudes where they form (e.g., Silber et al., 2017b).
4. Understanding the risks and further constraining the lower boundary of sizes, compositions and velocities of large objects that create shock waves in the lower atmosphere and represent a potential hazard (Mironov et al., 2015).
5. Understanding the hazards of meteor generated shock waves and their effects in MLT to future frequent space travel (Mironov and Murtazov, 2015).

**Acknowledgements**

The authors thank the Editors of ASR for inviting this review paper, and the two anonymous reviewers for their helpful comments to improve this paper. EAS gratefully acknowledges useful discussions with Reynold E. Silber. EAS also acknowledges the Natural Sciences and Engineering Research Council of Canada Postdoctoral Fellowship program for supporting this project. MG acknowledges support from the ERC Advanced Grant No. 320773, and the Russian Foundation for Basic Research, project nos. 16-05-00004, 16-07-01072, and 18-08-00074. Research at the Ural Federal University is supported by the Act 211 of the Government of the Russian Federation, agreement No 02.A03.21.0006.

**List of variables (Sections 2 – 6)**

| | | | |
|---|---|---|---|
| $a$ | accommodation coefficient | $T_0$ | temperature, ambient air |
| $A_s$ | the dimensionless shape factor for a specific meteoroid shape | $T_m$ | mean temperature, meteoroid |
| | | $T_s$ | temperature, meteoroid surface |
| $c$ | speed of sound | $t$ | time |
| $C$ | specific heat of the meteoroid | $v$ | meteoroid velocity (or velocity of the air stream over the meteoroid) |
| $C_D$ | drag coefficient | | |
| $d$ | diameter, blunt object | $\bar{v}_v$ | mean velocity of the reflected/evaporated atoms and molecules from the meteoroid surface, and ahead of the shock wave |
| $d_m$ | diameter, meteoroid | | |
| $E$ | energy | | |
| $E_{vapor}$ | energy, ablated vapor | $V_m$ | volume, meteoroid |
| $h$ | altitude | $V_1$ | volume, upstream flow |
| $k$ | Boltzmann constant | $V_2$ | volume, shock |
| $Kn$ | Knudsen number | $v_1$ | flow velocity, free stream |
| $Kn_v$ | Knudsen number within vapor cap | $v_2$ | flow velocity, shock |
| $m$ | mass, meteoroid | $v_{particles}$ | velocity, vapor cloud molecules |
| $M_\infty$ | Mach number | $x$ | distance traveled by a blunt body |
| $m_a$ | mass, air molecule | $\gamma$ | specific heat ratio |
| $m_m$ | mass, meteoric molecule/atom | $\delta$ | shock detachment distance |
| $M_{sw}$ | Mach number, shock wave | $\varepsilon$ | emissivity of the meteoroid |
| $N_v$ | the number of vaporizing meteoric molecules/atoms | $\Lambda$ | heat transfer coefficient (heat of ablation of the meteoroid material) |
| $p$ | pressure, shock | $\mu$ | dynamic viscosity coefficient |
| $p'$ | pressure behind compression shock | $\rho$ | density |
| $p_0$ | pressure, ambient atmosphere | $\rho'$ | density behind compression shock $\rho_0$ density, ambient atmosphere |
| $p_1$ | pressure, free stream | | |
| $p_2$ | pressure, shock | $\rho_1$ | density, free stream |
| $Q$ | latent heat of vaporization | $\rho_2$ | density, shock |
| $R_0$ | blast (or characteristic) radius | $\rho_a$ | density, air |
| $r_0$ | radius, initially formed meteor trail | $\rho_m$ | density, meteoroid |
| $Re$ | Reynolds number | $\sigma_a$ | ablation coefficient |
| $r_m$ | radius, meteoroid | $\sigma_{SB}$ | Stefan-Boltzmann constant |
| $S$ | projected cross-sectional area | $\tau_c$ | characteristic time, chemical reactions |
| $T$ | temperature | $\tau_f$ | characteristic time, for a fluid element to travel the distance of the flow field |
| $T'$ | temperature behind compression shock | | |
| | | $\Phi$ | Mach cone angle |

**List of variables (Section 7)**

| | | | |
|---|---|---|---|
| $h$ | enthalpy per unit mass | $T_2$ | temperature, shock layer |
| $h_0$ | total enthalpy per unit mass | $u, v, w$ | components of velocity $V$ |
| $h_1$ | enthalpy per unit mass, undisturbed flow | $U_i$ | velocity of the $i^{th}$ species in the flow field |
| $h_2$ | enthalpy per unit mass, shock | $V$ | velocity (vector form), flow field |
| $k$ | thermal conductivity | $v_1$ | velocity, undisturbed flow |
| $M_i$ | molar mass of species ($i$) | $v_2$ | velocity, shock layer |
| $p$ | pressure | $\lambda$ | bulk viscosity coefficient |
| $p_1$ | pressure, undisturbed flow | $\mu$ | dynamic viscosity coefficient |
| $p_2$ | pressure, shock layer | $\rho$ | density |
| $q_R$ | radiative heat | $\rho_1$ | density in undisturbed flow |
| $T$ | temperature | $\rho_2$ | density, shock layer |
| $T_1$ | temperature, undisturbed flow | | |